\documentclass[12pt]{article}

\usepackage{amsmath,amssymb,amsthm}

\usepackage{epsfig}
\usepackage{arydshln}

\newcommand{\bfm}[1]{\mbox{\boldmath ${#1}$}}
\newcommand{\nonum}{\nonumber \\}
\newcommand\eq[1] {(\ref{#1})} 
\newcommand{\beqa}{\begin{eqnarray}}
\newcommand{\eeqa}[1]{\label{#1}\end{eqnarray}}
\newcommand{\beq}{\begin{equation}}
\newcommand{\eeq}[1]{\label{#1}\end{equation}}
\newcommand{\Grad}{\nabla}
\newcommand{\Div}{\nabla \cdot}
\newcommand{\Curl}{\nabla \times}

\newcommand{\Real}{\mathop{\rm Re}\nolimits}
    \newcommand{\Imag}{\mathop{\rm Im}\nolimits}
\newcommand{\Tr}{\mathop{\rm Tr}\nolimits}

\newcommand{\lang}{\langle}
\newcommand{\rang}{\rangle}
\newcommand{\Md}{\partial}

\newcommand{\Ga}{\alpha}
\newcommand{\Gb}{\beta}

\newcommand{\Ge}{\epsilon}

\newcommand{\Gg}{\gamma}

\newcommand{\Gk}{\kappa}

\newcommand{\Gl}{\lambda}
\newcommand{\Gn}{\eta}
\newcommand{\Gm}{\mu}

\newcommand{\Gj}{\tau}

\newcommand{\Go}{\omega}
\newcommand{\Gx}{\xi}

\newcommand{\Gz}{\zeta}

\newcommand{\GG}{\Gamma}

\newcommand{\GT}{\Theta}

\newcommand{\GO}{\Omega}

\newcommand{\BGa}{\bfm\alpha}
\newcommand{\BGb}{\bfm\beta}

\newcommand{\BGe}{\bfm\epsilon}

\newcommand{\BGg}{\bfm\gamma}

\newcommand{\BGn}{\bfm\eta}
\newcommand{\BGm}{\bfm\mu}

\newcommand{\BGj}{\bfm\tau}

\newcommand{\BGY}{\bfm\Psi}

\newcommand{\CM}{{\cal M}}

\newcommand{\bpm}{\begin{pmatrix}}
\newcommand{\epm}{\end{pmatrix}}
\newcommand\fig[1] {{\rm Figure}~\ref{fig:#1}}

\newcommand\labfig[1] {\label{fig:#1}}
\newcommand\sect[1] {\ref{sect:#1}}
\newcommand\labsect[1] {\label{sect:#1}}

\def\Bb{{\bf b}}

\def\Bg{{\bf g}}
\def\Bh{{\bf h}}

\def\Bn{{\bf n}}

\def\Br{{\bf r}}

\def\Bu{{\bf u}}
\def\Bv{{\bf v}}

\def\Bx{{\bf x}}

\def\BF{{\bf F}}
\def\BG{{\bf G}}
\def\BH{{\bf H}}
\def\BI{{\bf I}}

\def\BL{{\bf L}}

\def\half{{\scriptstyle{1\over 2}}}

\usepackage{graphics}
\usepackage{subcaption}
\usepackage[dvipsnames]{xcolor}
\newcommand{\cb}{\color{black}}

\usepackage{url}
\usepackage{hyperref}

\title{Determining the volume fraction in 2-phase composites and bodies using time varying applied fields}
\author{Ornella Mattei$^1$, Graeme W. Milton$^2$, and Mihai Putinar$^3$}
\date{\small{$^1$ Department of Mathematics, San Francisco State University, CA 94132, USA,\\
$^2$Department of Mathematics, University of Utah, Salt Lake City, UT 84112, USA, \\
    $^3$Department of Mathematics, University of California at Santa Barbara, CA 93106, USA, 
    and School of Mathematics, Statistics and Physics, Newcastle University, NE1 7RU Newcastle upon Tyne, 
UK.
\\Emails: mattei@sfsu.edu, milton@math.utah.edu, mputinar@math.ucsb.edu, mihai.putinar@ncl.ac.uk}}

\begin{document}
\maketitle
\vspace{2ex}
\begin{abstract}
  A body $\GT$ containing two phases, which may form a periodic composite with microstructure much smaller that the body, or which may have structure on a length scale comparable to the body, is subjected to slowly
  time varying boundary conditions that
  would produce an approximate uniform field in $\GT$ were it filled with
  homogeneous material. Here slowly time varying means that the wavelengths and
  attenuation lengths of waves at the frequencies associated with the time
  variation are much larger \cb than \cb  the size of $\GT$, so that we can make a
  quasistatic approximation. At least one of the two phase does
  not have an instantaneous response but rather depends on fields at
  prior times. The fields may be those associated with electricity, magnetism,
  fluid flow in porous media, or antiplane elasticity. We find, subject
  to these approximations, that the time variation of the boundary conditions
  can be designed so boundary measurements at a specific time $t=t_0$
  exactly yield the volume fractions of the phases, independent of the detailed
  geometric configuration of the phases. Moreover, for specially tailored
  time variations, the volume fraction can be exactly determined from
  measurements at any time $t$, not just at the specific time $t=t_0$. We also
  show how time varying boundary conditions, not oscillating at the single
  frequency $\Go_0$, can be designed to exactly retrieve the response at $\Go_0$.

\end{abstract}
\vspace{3ex}
\section{Introduction}
\setcounter{equation}{0}
\labsect{0}

Consider a body $\GT$ containing a periodic composite material of two isotropic phases with the cell size
being much smaller than $\GT$. We ask:
can one exactly determine the volume fractions of the two phases
from the response of the body to time varying applied electric, magnetic, or elastic quasistatic fields? Here
quasistatic means that the frequencies associated with the time variation have waves with wavelengths
and attenuation lengths much larger than $\GT$.

Cherkaev \cite{Cherkaev:2001:IHE} realized the answer is yes, that in principle one can recover the volume fraction exactly. One makes
measurements of, say, the electrical response, as governed by the effective electrical permittivity $\BGe_*(\Go)$ at each of a continuum of
frequencies $\Go$, such that the
ratio $\Ge_1(\Go)/\Ge_2(\Go)$ of the permittivities  $\Ge_1(\Go)$, and $\Ge_2(\Go)$ of the phases traces an arc in
the complex plane. In principle this allows one by analytic continuation to determine the function $\BGe_*(\Ge_1,\Ge_2)$
and hence to obtain the measure entering the Stieltjes function
representation of $\BGe_*/\Ge_2$ as a function of  $\Ge_1/\Ge_2$. Then, the integral of the measure determines the volume
fraction. Besides the difficulty of measuring the response at sufficiently many frequencies that approximate a continuum of frequencies,
the analytic continuation is ill-posed. Nevertheless, one can get approximations to the measure and thus to the volume fraction. In prior work \cite{McPhedran:1982:ESI, Day:2000:SFC}
and subsequent work \cite{Zhang:2008:PAI, Cherkaev:2008:DRM, Cherkaev:2011:CSP} this was done either by estimation of the distributions of poles and zeros or poles and residues
when the measure is discrete or approximated by a discrete one, or by extraction of a continuous measure. 
Other work based on the analytic properties of $\BGe_*(\Ge_1,\Ge_2)$ include using in an inverse way \cite{McPhedran:1982:ESI, McPhedran:1990:ITP, Cherkaeva:1998:IBM, Engstrom:2005:BET}
volume fraction dependent bounds on $\BGe_*(\Ge_1(\Go),\Ge_2(\Go))$ at one frequency $\Go$ or correlating the values of  $\BGe_*(\Ge_1(\Go_i),\Ge_2(\Go_i))$
at $n$ frequencies $\Go_i, i=1,2,\ldots,n$. To yield accurate approximations to the volume fraction these approaches usually require measurements
at many frequencies.


Here we show how the volume fractions can be exactly obtained from a single time varying quasistatic field, composed
of a continuum of frequencies. By carefully tailoring the time varying applied field, the response at a selected
time $t_0$ only depends on the volume fractions and not on the detailed microstructure. To our surprise, in many cases we
find that the response at any time $t$, and not just $t_0$, only depends on the volume fractions and not on the detailed microstructure.

Our work is an extension of that in \cite{Mattei:2016:BRL, Mattei:2016:BRV, Mattei:2020:EPA}. In \cite{Mattei:2016:BRL, Mattei:2016:BRV}  bounds were obtained for antiplane elasticity
on the response of the average stress $\lang\BGj\rang(t)$, given a time dependent average strain
$\lang\BGe\rang(t)=\BGe_0H(t)$, where the angular brackets denote a volume average over the cell of periodicity,
$\BGe_0$ is a fixed vector and $H(t)$ is the Heaviside function, $0$ for $t<0$ and
$1$ for $t>0$. The bounds were volume fraction dependent, but otherwise independent of 
the microstructure (and some bounds were also independent of the volume fraction).
Remarkably, in some examples the bounds were exceedingly tight at particular times. These bounds can be used in an
inverse way to provide tight bounds on the volume fraction given measurements of the response  $\lang\BGj\rang(t)$ at these particular
times.

For some choices of the frequency dependent shear moduli $\Gm_1(\Go)$ and $\Gm_2(\Go)$ of phase 1 and phase 2
the bounds in \cite{Mattei:2016:BRL, Mattei:2016:BRV} were quite wide at all times. While the tightness or lack
of tightness of the bounds was
explained, it was unclear how to tailor $\lang\BGe\rang(t)$ to get
tight bounds at a given time $t=t_0$, and whether this was at all possible when
the bounds with $\lang\BGe\rang(t)=\BGe_0H(t)$ were not tight. The case of applied fields that
were a finite sum of $n$ terms with a $e^{-i\Go_jt}$ time dependence, with complex frequencies
$\Go_1,\Go_2,\ldots,\Go_n$ was studied in \cite{Mattei:2020:EPA}. To ensure that the input function
did not diverge to infinity in the distant past, we required that $\Imag\Go_j\geq 0$ for all $j$.
Designing these input signals to ensure tight bounds on the response at time $t=t_0$ boiled
down to one of approximation on the real interval $[-1,1]$ of a polynomial of degree $m+1$
by one of degree $m$, achieved by letting the difference be an appropriately scaled
Chebyshev polynomial. Additionally, we found the input function could be designed to
approximately reproduce at time $t_0$ the response at another frequency $\Go_0$. 

Our approach here uses judicious applications of the residue theorem to recover the volume fraction exactly.
Moreover we can extract additional information such as the exact first moment of the measure and
the exact response at a frequency $\Go_0$. We emphasize that is important for the wavelengths and attenuation lengths
of the frequencies associated with the time variation of the applied field be much bigger that $\GT$, and not just
the microstructure. Otherwise, the frequency components will dephase or attenuate differently as one moves inside $\GT$.  

Exact values of the volume fractions can sometimes be obtained using another
approach. It has an entirely different character. 
While the macroscopic response of composite materials, as governed by
their effective moduli, is generally microstructure dependent, 
there is a plethora of examples where
effective moduli satisfy microstructure independent relations, or relations that
only involve the volume fractions of the phases.
A classic example is Hill's
formula for the effective Lame modulus $\Gl_*$ of a mixture of two elastic phases
sharing the same shear modulus $\Gm$, having Lame moduli $\Gl_1$ and $\Gl_2$,
and occupying volume fractions $f_1$ and $f_2=1-f_1$:
\beq \frac{1}{\Gl_*+2\mu}=\frac{f_1}{\Gl_1+2\mu}+\frac{f_2}{\Gl_2+2\mu}. \eeq{0.14}
So if $\Gl_*$ and the moduli of the phases are known, we can determine the volume fractions \cb
$f_1$ and $f_2$ exactly, \cb without having to know anything about the detailed microstructure. Many
examples of such exact relations are surveyed in chapters 3,4,5, and 6 in \cite{Milton:2002:TOC}.
A general theory of exact relations was developed \cite{Grabovsky:1998:EREa,Grabovsky:2000:ERE}
and was reviewed in chapter 17 of \cite{Milton:2002:TOC} and in \cite{Grabovsky:2016:CMM}.
It led to an enormous number of new exact relations \cite{Grabovsky:2016:CMM}. 
According to this theory,  a manifold $\CM$ is an exact relation if whenever a material has a tensor
field $\BL(\Bx)\in\CM$ for all $\Bx$ then its associated effective tensor $\BL_*$ also lies in $\CM$. 
Under an appropriate fractional linear transformation, $\CM$ is transformed to a
linear subspace that must satisfy certain algebraic properties. The theory is easily extended to include volume fractions
by letting $\CM$ consist of pairs $(f,\BL)$, where $f$ represents the volume fraction, or volume fractions, of the phases.

Our results are also applicable to recovering the volume and shape of an inclusion, or inclusions, in the body $\GT$ from exterior boundary measurements. This is an important inverse problem
that has received much attention:
see \cite{Calderon:1980:IBV, Kohn:1984:DCB, Kohn:1987:RVM, Sylvester:1987:GUT, Sylvester:1993:LAI, Kang:1997:ICP, Alessandrini:1998:ICP, Ikehata:1998:SEI, Bruhl:2003:DIT,
  Capdeboscq:2003:OAE, Ammari:2004:RSI, Ammari:2007:PMT,
  Kirsch:2011:SOH, Mueller:2012:LNI, Kolokolnikov:2015:RMS} and references therein.
For the extension of our work to this problem the reader can refer to Sections 8,9, and 10 of
\cite{Mattei:2020:EPA}.
The generalization to recovering the volume of an inclusion, or inclusions, in the body $\GT$ from exterior boundary measurements
is straightforward and will be outlined in Section \sect{XX}. 

\section{Preliminaries}
\setcounter{equation}{0}
\labsect{0a}

The effective magnetic permeability tensor (matrix) $\BGm_*$ of a periodic composite containing two
phases with isotropic permeabilities $\Gm_1$ and $\Gm_2$ is given by solving
the equations
\beq \Div\Bb=0,\quad \Bh=-\Grad\psi,\quad\Bb(\Bx)=\Gm(\Bx)\Bh(\Bx), \eeq{0.1}
where $\Bb$ is the magnetic induction, $\Bh$ the magnetic field, $\psi$ the magnetic scalar potential and 
the permeability $\Gm(\Bx)$ takes the value $\Gm_1$ in phase 1
and $\Gm_2$ in phase 2, and $\Gm(\Bx)$, $\Bh(\Bx)$, and $\Bb(\Bx)$ are all periodic
functions of $\Bx$. Let $\lang\cdot\rang$ denote an average over the unit cell $Q$ of
periodicity:
\beq \lang f\rang=\frac{1}{|Q|}\int_Q f(\Bx)\,d\Bx, \eeq{0.1a}
where $|Q|$ denotes the volume of $Q$. 
As the response field $\lang\Bb\rang$ depends linearly on the applied field $\lang\Bh\rang$,
we may write
\beq \lang\Bb\rang=\BGm_*\lang\Bh\rang,
\eeq{0.2}
which thus determines the effective permeability tensor $\BGm_*$.

We could equivalently deal with electricity, fluid flow in porous media, and antiplane elasticity
as they are governed by the same equations, with the electric permittivity, fluid permeability, and
shear modulus playing the role of the permeability. For example, in antiplane elasticity in an isotropic material one assumes
that the shear modulus $\Gm$, which takes the role of the magnetic permeability,
is independent of $x_3$, $\Gm=\Gm(x_1,x_2)$, and one looks for a solution where
the only non-zero displacement component is $u_3$ and that is independent of $x_3$. Then the
only non-zero components of the strain $\BGe(\Bx)=[\Grad\Bu+(\Grad\Bu)^T]/2$ are
$\Ge_{13}=\Ge_{31}=\half\Md u_3/\Md x_1$ and $\Ge_{23}=\Ge_{32}=\half\Md u_3/\Md x_2$ . \cb Because \cb $\BGe(\Bx)$ is a pure shear (trace-free),
the stress is $\BGj(x_1,x_2)=2\Gm(x_1,x_2)\BGe(x_1,x_2)$ and its only non-zero components are
\beq \Gj_{13}=\Gj_{31}=2\Gm\Ge_{13}=\Gm\Md u_3/\Md x_1, \quad \Gj_{23}=\Gj_{32}=2\Gm\Ge_{23}=\Gm \Md u_3/\Md x_2.
\eeq{0.8}
\cb Since \cb $\Div\BGj=0$, we get $\Md\Gj_{13}/\Md x_1+\Md\Gj_{23}/\Md x_2=0$. Thus, the two-dimensional version
of the equations \eq{0.1} hold with the replacements
\beq \Bb\to \bpm \Gj_{13} \cr \Gj_{23} \epm,\quad \Bh\to \bpm \Md u_3/\Md x_1 \cr \Md u_3/\Md x_2 \epm, \quad \Gm(\Bx)\to \Gm(x_1,x_2). 
\eeq{0.9}
\subsection{Analytic properties}

The effective permeability
as $\BGm_*$ is an analytic function of $\Gm_1$ and $\Gm_2$, except at negative real
values of the ratio $\Gm_1/\Gm_2$ with the properties:
\beqa
\Imag[\BGm_*(\Gm_1,\Gm_2)]& > & 0\text{ if }\Imag\Gm_1>0\text{ and }\Imag\Gm_2>0, \nonum
\BGm_*(\Ga\Gm_1,\Ga\Gm_2)& = & \Ga\BGm_*(\Gm_1,\Gm_2), \quad \BGm_*(1,1)=\BI, \nonum
 \frac{\Md\BGm_*(\Gm_1,1)}{\Md\Gm_1}\Biggl{|}_{\Gm_1=1} & = &f_1\BI,\quad
\Tr\frac{\Md^2\BGm_*(\Gm_1,1)}{\Md\Gm_1^2}\Biggl{|}_{\Gm_1=1}=-2f_1f_2 ,
\eeqa{0.3}
where $\Tr$ denotes the trace of the matrix.
Taking $\Ga=i$ yields the corollary that
\beq \Real[\BGm_*(\Gm_1,\Gm_2)] >  0\text{ if }\Real\Gm_1>0\text{ and }\Real\Gm_2>0.
\eeq{0.4}
As a result of these analytic properties $\BGm_*(\Gm_1,\Gm_2)$ has the integral representation
\cite{Bergman:1978:DCC, Milton:1981:BCP, Golden:1983:BEP}
\beq \BGm_*(\Gm_1,\Gm_2)=\Gm_2\BI+2f_1\Gm_2\BG_\Gn(z), \text{        with } z=\frac{\Gm_1+\Gm_2}{\Gm_2-\Gm_1}
\eeq{0.5}
 \cb where
\beq \BG_\Gn(z)=\int_{-1}^1\frac{d\BGn(\Gl)}{\Gl-z}, \eeq{0.6}
 is a Markov function of $z$.
$\BG_\Gn(z)$ has positive-semidefinite matrix valued measure $\BGn$ \cb dependent on the material microstructure and having the identity as its mass,
\beq \int_{-1}^1\,d\BGn(\Gl)=\BI.
\eeq{0.7}
\cb Furthermore,  as implied by \eq{0.3}, the first moment of the measure satisfies \cb
\beq \Tr\int_{-1}^1\Gl\,d\BGn(\Gl)=2f_2-d, \eeq{0.7a}
where $d=\Tr\BI$ is the dimensionality, $d=2$ or $3$. In the case where the composite has square (for $d=2$) or cubic
symmetry (for $d=3$), or is isotropic, then $d\BGn(\Gl)$ is proportional to the identity matrix and \eq{0.7a} implies
\beq \int_{-1}^1\Gl\,d\BGn(\Gl)=(2f_2/d-1)\BI. \eeq{0.7b}

Notice from \eq{0.5} that $f_1$ can be absorbed into
the measure, in which case the integral of the measure yields the volume fraction.

\subsection{Time dependent fields}

If there is some
time dependence, then in general the \cb magnetic induction \cb at a given point and time depends
on the \cb magnetic field \cb at that point at previous times (memory effect). Thus, the
last relation in \eq{0.1}, called the constitutive equation, gets replaced by
\beq \Bb(\Bx,t)=\int_{-\infty}^t K(\Bx,t-\tilde{t})\Bh(\Bx,\tilde{t})\,d\tilde{t}, \eeq{0.10}
where by causality $K(\Bx,s)=0$ if $s<0$.
Physically $\Bh(\Bx,t)$ is always real but it is useful to also consider complex fields. In particular, we may consider fields $\Bh(\Bx,t)$ with a
time dependence $e^{-i\Go t}$ where the complex frequency $\Go$ has positive imaginary part, to ensure that the applied field
is small in the distant past. Then, \eq{0.10}, with $\tilde{t}$ replaced by $t-s$, \cb implies \cb
\beq \Bb(\Bx,t)=\Gm(\Bx,\Go)\Bh(\Bx,t), \quad \Gm(\Bx,\Go)=\int_0^{\infty} K(\Bx,s)e^{i\Go s}\,ds
\eeq{0.11}
Thus, the constitutive relation in \eq{0.1} still holds, but with complex fields and complex permeabilities.
Since  $K(\Bx,s)$ is real, the complex permeability $\Gm(\Bx,\Go)$ satisfies
\beq \overline{\Gm(\Bx,\Go)}=\Gm(\Bx,-\overline{\Go}). \eeq{0.12}
The constraints
that $\Div\Bb=0$ and $\Curl\Bh=0$ are valid in the quasistatic limit, where the wavelength is much larger than the size of the body $\GT$. We can take applied fields $\lang\Bh\rang(t)$ that are a superposition $\Bu(t)$ of a continuum of complex frequencies
\beq  \lang\Bh\rang(t)=\Bu(t)=\int_{0}^1\BGb(s) e^{-i\Go(s)(t-t_0)}\,ds, \eeq{1.12a}
where we are free to choose the amplitude vector field $\BGb(s)$. To ensure
that \eq{1.12a} approximately holds for all unit cells $Q$ within $\GT$,
we need to impose
boundary conditions that would ensure uniformity of the field in $\GT$ were it
filled with a homogeneous medium: these are affine Dirichlet boundary conditions
on $\psi$ at $\Md\GT$: see Section \sect{XX}. 

The associated field $\lang\Bb\rang(t)$ in the case when
the composite is entirely phase 2, is
\beq \lang\Bb\rang_2(t)= \int_{0}^1 \Gm_2(\Go(s))\BGb(s)e^{-i\Go(s)(t-t_0)}\,ds.
\eeq{0.12aa}
The corresponding output field can be considered to be
\beq \Bv(t)=\lang\Bb\rang(t)-\lang\Bb\rang_2(t)
= \int_{0}^1 2f_1\Gm_2(\Go(s))\BG_\Gn(z(\Go(s)))\BGb(s)e^{-i\Go(s)(t-t_0)}\,ds,
\eeq{0.13}
with
\beq
z(\Go)=\frac{\Gm_1(\Go)+\Gm_2(\Go)}{\Gm_2(\Go)-\Gm_1(\Go)},
\eeq{0.13a}
The physical applied field will be $\Real\lang\Bh\rang(t)$, and the physical output field will be
\beq \Real \Bv(t)=\Real\lang\Bb\rang(t)-\Real\lang\Bb\rang_2(t). 
\eeq{0.13b}
\subsection{The dual problem}
Alternatively, one may choose $\Bu(t)$ to be the average magnetic induction $\lang\Bh(t)\rang$, and the constitutive and effective
constitutive law can be rewritten as
\beq \Bh(\Bx)=\Gm(\Bx)^{-1}\Bb(\Bx),\quad \lang\Bh\rang=\BGm_*^{-1}\lang\Bb\rang. \eeq{p.1}
Rather than considering the function $\BGm_*(\Gm_1,\Gm_2)$ we can consider $\BGm_*^{-1}$ as a function $\BGm_*^{-1}(1/\Gm_1,1/\Gm_2)$ of $1/\Gm_1$ and $1/\Gm_2$.
All but the last property in \eq{0.3} hold if we make the replacements
\beq \Gm_1\to 1/\Gm_1,\quad\Gm_2\to 1/\Gm_2, \quad \BGm_* \to \BGm_*^{-1}. \eeq{p.2}
Consequently $\BGm_*^{-1}(1/\Gm_1,1/\Gm_2)$ has the integral representation
\beq \BGm_*^{-1}(1/\Gm_1,1/\Gm_2)=\BI/\Gm_2+2f_1\BH_\Gj(z)/\Gm_2,\text{  with  }z=-\frac{1/\Gm_1+1/\Gm_2}{1/\Gm_2-1/\Gm_1}=
\frac{\Gm_1+\Gm_2}{\Gm_2-\Gm_1},
\eeq{p.3}
where $\BH_\Gj(z)$ is another Markov function of $z$,
\beq \BH_\Gj(z)=\int_{-1}^1\frac{d\BGj(\Gl)}{\Gl-z}, \eeq{p.4}
with positive-semidefinite matrix valued measure $\BGj$ dependent on the material microstructure and having the identity as its mass.
With $\Bu(t)$ as in \eq{1.12a} (but with  $\lang\Bh\rang(t)$ replaced by  $\lang\Bb\rang(t)$), we can take our output field to be
\beq  \Bv(t)=\lang\Bh\rang(t)-\lang\Bh\rang_2(t)
= \int_{0}^1 \frac{2f_1\BH_\Gj(z(\Go(s)))}{\Gm_2(\Go(s))}\BGb(s)e^{-i\Go(s)(t-t_0)}\,ds,
\eeq{p.5}
where 
\beq \lang\Bh\rang_2(t)= \int_{0}^1 \BGb(s)e^{-i\Go(s)(t-t_0)}/\Gm_2(\Go(s))\,ds.
\eeq{p.6}
is the field $\lang\Bh\rang(t)$ in the case when the composite is entirely phase 2, and $z(\Go)$ is given by \eq{0.13a}.
The analysis proceeds in a parallel way. Again, to ensure that
$\lang\Bb\rang(t)$ is almost independent of  \cb the unit cell of periodicity \cb $Q$, we need to impose Neumann
boundary conditions on $\Bb(\Bx)$ at $\Md\GT$ that would ensure
uniformity of the field in $\GT$ were it filled with a homogeneous medium:
see Section \sect{XX}.

\subsection{Determining the volume occupied by inclusions in a body}
\labsect{XX}
Rather than assuming $\GT$ contains a microstructured material, we can treat the case where
$\GT$ contains two phases, with structure not necessarily small compared to $\GT$. For instance
$\GT$ may contain inclusions of one phase surrounded by the second phase, where
the sizes of the inclusions may be comparable to the size of $\GT$. The volume fraction $f_1$, or $f_2=1-f_1$,
is now the volume occupied by phase 1, or phase 2, divided by $|\GT|$, the volume of $\GT$ which may be
presumed to be known.

We redefine our average as
\beq \lang f\rang=\frac{1}{|\GT|}\int_\GT f(\Bx)\,d\Bx, \eeq{y.1}
where $|\GT|$ is the volume of $\GT$. We choose affine Dirichlet boundary
conditions on the potential $\psi$ (which may make more sense in an electrical setting or in the setting of antiplane \cb elasticity\cb)
that are dictated by our input signal $\Bu(t)$:
\beq \psi(\Bx,t)=-\Bu(t)\cdot\Bx\text{ on }\Md\GT. \eeq{y.2}
Equivalently, as more appropriate to magnetism,
one can specify the tangential components of $\Bh$ at $\Md\GT$. As $\Bh=-\Grad\psi$
a straightforward calculation shows that $\lang \Bh\rang(t)=\Bu(t)$. 
Since $\Div\Bb=0$ it then follows that
\beq \lang \Bb\rang=  \lang \Div(\Bb\Bx)\rang=\frac{1}{|\GT|}\int_{\Md\GT}(\Bb\cdot\Bn)\Bx~dS,
\eeq{y.3}
where $\Bn$ is the outwards normal to $\Md\GT$, {and  $\Bb\Bx$ is the outer product between the vectors $\Bb$ and $\Bx$}.
So \cb $\lang \Bb\rang$ \cb can be determined by measurements at the boundary of $\GT$. The relation between $\lang \Bb\rang$ and
$\lang \Bh\rang$ defines some sort of effective tensor $\BGm_*^D$ ($D$ for Dirichlet):
\beq \lang \Bb\rang=\BGm_*^D\lang \Bh\rang, \eeq{y.4}
and the function $\BGm_*^D(\Gm_1,\Gm_2)$ has the same analytic properties as $\BGm_*(\Gm_1,\Gm_2)$ in \eq{0.3}, excepting the last property,
and can be represented in terms of an associated non-negative measure $d\BGn^D(\Gl)$. Therefore the analysis carries
over \cb with \cb $\Bu(t)=\lang \Bh\rang(t)$ given by \eq{1.12a}.

We can also consider the dual problem where we specify Neumann boundary conditions on $\Bb$:
\beq \Bb\cdot\Bn=\Bu(t)\cdot\Bn\text{ on }\Md\GT, \eeq{y.5}
which from \eq{y.3} implies $\lang \Bb\rang(t)=\lang \Bu(t)\rang=\Bu(t)$. We can measure
\beq \lang\Bh\rang=\lang -\Grad\psi \rang= -\frac{1}{|\GT|}\int_{\Md\GT}\psi\Bn~dS,
\eeq{y.6}
and the relation between $\lang \Bh\rang$ and
$\lang \Bb\rang$ defines some other sort of effective tensor $\BGm_*^N$ ($N$ for Neumann):
\beq \lang \Bh\rang=(\BGm_*^N)^{-1}\lang \Bb\rang. \eeq{y.7}
Again $(\BGm_*^N)^{-1}$ as a function of $1/\Gm_1$ and $1/\Gm_2$ has the same analytic properties as $\BGm_*(\Gm_1,\Gm_2)$ in \eq{0.3}, excepting the last property,
and can be represented in terms of an non-negative associated measure $d\BGj^N(\Gl)$.

If the body does contain a periodic composite material of two isotropic phases with the cell size being much smaller than $\GT$, then we may equate
the Dirichlet and Neumann effective tensors,
\beq \BGm_*^D=\BGm_*^N, \eeq{y.8}
as the boundary conditions \eq{y.5} will produce fields $\Bh$ and $\Bb$ that are almost periodic and consequently
\eq{y.2} will be satisfied in the homogenized limit.

\section{Choosing an input signal that allows the volume fraction to be exactly determined}
\setcounter{equation}{0}
\labsect{1}
\cb In the previous sections we introduced a general framework to handle both the problem of determining the volume fractions in a periodic composite and that of determining the volume occupied by inclusions in a body, by applying affine boundary conditions of either the Dirichlet or Neumann type. Within such a framework, the input signal (that is $\lang\Bh\rang(t)$, or $\lang\Bb\rang(t)$ for the dual problem) is
\beq \Bu(t)=\int_{0}^1\BGb(s) e^{-i\Go(s)(t-t_0)}\,ds, \eeq{1.1}
where the $\Go(s)$ parameterizes a curve in the upper half complex plane,
and the vector valued function $\BGb(s)$ needs to be determined so that the volume fraction
is determined at least at the given time $t_0$, {as showed in the remaining of this section}. The output 
is (see \eqref{0.13}, and \eqref{p.5} for the dual problem)
\beq  \Bv(t)=\int_{0}^1 a_0 \BF_\Gg (z(\omega(s)))\BGa(s)e^{-i\Go(s)(t-t_0)}\,ds,
\eeq{1.2}
in which $\BF_\Gg(z)$ is given by either $\BG_\Gn(z)$  \eqref{0.6} or $\BH_\Gj(z)$ \eqref{p.4}, 
and
\beq \BGa(s)=\BGb(s) c(\Go(s)), \eeq{1.5}
where the functions  $z(\omega)$ and $c(\omega)$ depend on $\Go$ in some known way, with
\beq z(\Go)=\overline{z(-\overline{\Go})}.
\eeq{1.5a}
Specifically, $ z(\Go)$ takes the same expression for both problems (see eqs. \eqref{0.13a} and \eqref{p.3}), whereas $c(\Go)=\Gm_2(\Go)$ for the direct problem and $c(\Go)=1/\Gm_2(\Go)$ for the dual problem. 
The real constant $a_0=2f_1$ and the unknown measure $d\BGg$, corresponding to either $d\BGn$ or $d\BGj$, depend on the system.
{Note that, due to equation \eqref{1.5}, choosing the function $\BGb(s)$ is equivalent to choosing the function $\BGa(s)$, since  $c(\Go(s))$ is known.}

As it is only the real part of  $\Bv(t)$ that has physical significance, we can write
\beq \Real \Bv(t_0)=a_0\int_{-1}^1 d\BGg(\Gl) \Bg(\Gl),
\eeq{1.6}
\cb
where
\beqa \Bg(\Gl)& = &\frac{1}{2}\int_{0}^1\frac{\BGa(s)}{\Gl-z(\Go(s))}\,ds
+\frac{1}{2}\int_{0}^1\frac{\overline{\BGa(s)}}{\Gl-\overline{z(\Go(s))}}\,ds \nonum
& = & \int_{C\cup\overline{C}}\frac{\Br(z)}{\Gl-z}\,dz,
\eeqa{1.7}
in which $C$ is the curve traced out by $z(\Go(s))$ as $s$ {\it increases} from $0$ to $1$,
$\overline{C}$  is the curve traced out by $\overline{z(\Go(s))}$ as $s$ {\it decreases} from $1$ to $0$,
and for $z\in C$, $\Br(z)$ is obtained from
\cb
\beq \Br(z(\Go(s)))=\BGa(s)/(2z'(\Go(s))\Go'(s)), \eeq{1.8}
\cb
while for $z\in \overline{C}$,
\beq \Br(z)=-\overline{\Br(\overline{z})}. \eeq{1.9}

Let us assume $C\cup\overline{C}$ is a closed curve encircling the interval $[-1,1]$ once anticlockwise.
We choose $\BGa(s)$ so that $\Br(z)$ satisfies \eq{1.9} and is analytic, with no poles, in $C\cup\overline{C}$.
Then,{this allows one to apply the residue theorem in equation \eqref{1.7} to obtain}
\beq \Bg(\Gl)=2\pi i\Br(\Gl).\eeq{1.10}
{The corresponding output at time $t_0$, see equation \eqref{1.6},  would then be}
\beq \Real \Bv(t_0)=2\pi i a_0\int_{-1}^1 d\BGg(\Gl) \Br(\Gl).
\eeq{output_r_spec}
{In order for the output to depend only on $a_0$ and eventually the first moment of the measure, $\BGa(s)$ and, therefore $\Br(z)$, by equation \eqref{1.8}, have to be suitably chosen. For example, if one takes }
\beq \Br(z)=-i(1+kz)\Bn_0/(2\pi), \eeq{1.10a}
where $\Bn_0$ is a constant real unit vector \cb and $k$ is a real coefficient\cb, we obtain
\beq \Real \Bv(t_0)=a_0\int_{-1}^1(1+k\Gl)d\BGg(\Gl)\Bn_0=a_0\Bn_0+a_0k\int_{-1}^1\Gl d\BGg(\Gl)\Bn_0. \eeq{1.11}
Thus, with this choice of $\Br(z)$ the fact that $\Real \Bv(t_0)$ only depends on $a_0$ and the
first moment of the measure is simply a consequence of the residue theorem.
{Note that there is considerable freedom in the choice of $\Br(z)$, still
leading to \eq{1.11}. Indeed, for example, to $\Br(z)$ we may add $A(1/z)\Bn_0$ where $A$
is an entire function, zero at the origin. A deformation of the contour path
 $C\cup\overline{C}$ to infinity shows that this does not affect the output.}

\section{Conditions that allow one to generate appropriate trajectories}
\setcounter{equation}{0}
\labsect{2}
To apply the results of the previous section we need to choose $\Go(s)$ so that $C\cup\overline{C}$ is a closed curve encircling the interval $[-1,1]$ once.
Recall that $z(\Go)$ takes the same expression for both problems, see \eqref{0.13a} and \eq{p.3}, 
where $\Gm_i(\Go)$, $i=1,2$, are the responses of the two phases satisfying
\beq \Gm_i(\Go)=\overline{\Gm_i(-\overline{\Go})}, \eeq{2.1a}
with
\beq \Imag(\Gm_i(\Go))\geq 0 \text{ in the quadrant }\Real \Go>0, \Imag \Go>0.
\eeq{2.2}
To begin let us suppose that $\Gm_2(\Go)=\Gm_2$ a positive \cb real \cb constant independent of $\Go$. As
\beq  z(\Go)=\frac{2\Gm_2}{\Gm_2-\Gm_1(\Go)}-1, \eeq{2.3}
we see that $\Imag z(\Go)>0$ in the quadrant $\Real \Go>0$, $\Imag \Go>0$. The identity
\eq{2.1a} implies that $z(\Go)$ is real on the positive imaginary axis. $\Gm_1(\Go)$
and hence $z(\Go)$ might also be real along intervals of the real $\Go$ axis.

Assume $z(\Go_1)$ and $z(\Go_2)$ are real at frequencies $\Go_1$ and $\Go_2$ each either on the
positive imaginary axis or positive real axis, with \cb  $z(\Go_1)> 1$ and $z(\Go_2)<-1$. \cb This is typically
guaranteed if \cb$\Gm_1(\Go_*)=\Gm_2$ \cb for some \cb$\Go_*$ \cb on the positive imaginary or real axis since
$z(\Go)$ will then have a simple pole at \cb $\Go=\Go_*$. \cb We may then take $\Go_1$ and $\Go_2$ on opposite
sides of \cb$\Go_*$ \cb and sufficiently close to \cb$\Go_*$\cb. We take a trajectory $\GG$
linking $\Go_1$ and $\Go_2$ that remains in the quadrant $\Real \Go>0$, $\Imag \Go>0$ except at the endpoints.
Its image is a curve $C=z(\GG)$ linking $z(\Go_1)$ and $z(\Go_2)$ that remains in
the upper half plane $\Imag z>0$. \cb Thus, $C\cup\overline{C}$ is a closed curve encircling the interval $[-1,1]$ once, anticlockwise.
The curve $C$ may have loops that do not cross the real axis. Alternatively, if  $\Gm_1(\Go)=\Gm_1$ is a positive  real  constant
independent of $\Go$, while  $\Gm_2(\Go)$ depends on $\Go$, then the curve $C=z(\GG)$ linking $z(\Go_1)$ and $z(\Go_2)$ will remain in
the lower half plane $\Imag z<0$. Thus $C\cup\overline{C}$ will be a closed curve encircling the interval $[-1,1]$ once, clockwise.
The latter case is treated by a simple modification of the arguments in the previous
section. \cb

More generally, if $\Gm_1(\Go)$ and $\Gm_2(\Go)$ both depend on $\Go$, we may 
consider those curves in the quadrant $\Real \Go>0$, $\Imag \Go>0$ where $\Gm_1(\Go)/\Gm_2(\Go)$, and hence $z(\Go)$
is real, and look for a point  $\Go_1$ on one of these curves \cb where $z(\Go_1)> 1$ and another point $\Go_2$ on the same or nearest
neighboring curve where $z(\Go_2)<-1$. \cb Again we join  $\Go_1$ and $\Go_2$ by a trajectory $\GG$ that avoids these curves.
This trajectory has an image curve $C=z(\GG)$ that lies in the upper
half-plane or in the lower half plane. 

In particular, if there is a $\Go_0$ in the quadrant
$\Real \Go_0>0$, $\Imag \Go_0>0$ such that  $\Gm_1(\Go_0)=\Gm_2(\Go_0)$ then there will be a curve in the $\Go$ plane (not a smooth
curve if $\Go_0$ is not a simple zero of $\Gm_1(\Go)-\Gm_2(\Go)$) along which $\Gm_1(\Go)/\Gm_2(\Go)$ is real
and $z(\Go)$ takes real values \cb $z(\Go_1)\gg 1$ and $z(\Go_2)\ll -1$ \cb at points $\Go_1$ and $\Go_2$
on this curve close to $\Go_0$. Then, we join  $\Go_1$ and $\Go_2$ by a trajectory $\GG$ that avoids the
curves where  $\Gm_1(\Go)/\Gm_2(\Go)$ is real.

              
\section{Recovering the response at another frequency}
\setcounter{equation}{0}
\labsect{3}
\cb
Suppose that we want to determine the response of the material at the frequency $\Go_0$:
\beq \Bv_0(t)=a_0\int_{-1}^1\frac{d\BGg(\Gl)\Bn_0}{\Gl-z_0}e^{-i\omega_0(t-t_0)}, \eeq{3.0}
where both the real constant $a_0$ and $z_0=z(\Go_0)$ are known, as well as the constant real unit vector $\mathbf{n}_0$. {Specifically, assume that  we are interested in the output function at time $t_0$:
\beq \Bv_0(t_0)=a_0\int_{-1}^1\frac{d\BGg(\Gl)\Bn_0}{\Gl-z_0}.\eeq{3.1}
Suppose that probing the material with an input function at the same frequency $\omega_0$ is not  feasible,  but it is possible to apply an input with a continuous spectrum of frequencies, such as the one in equation \eqref{1.1}. The corresponding output will then be \eqref{1.2}, and the goal of this section is to determine the function \(\boldsymbol{\beta}(s)\) in \eqref{1.1} so that the value taken by $\Bv(t)$ in \eq{1.2} at $t=t_0$ will be equal to \(\Bv_0(t_0)\).  We start by assuming, as before, that the curve $C$ traced out by $z(\omega(s))$ is such that $C\cup\overline{C}$ is a closed curve encircling the interval $[-1,1]$ once, anticlockwise. Then, the following choice of  $\Br(z)$ given by \eq{1.8} (note that $\BGa(s)$ in \eq{1.8} is related to $\BGb(s)$ by \eq{1.5} so that, as before, prescribing $\Br(z)$  is equivalent to choosing $\BGb(s)$):
\beq \Br(z)=\frac{i\Bn_0}{4\pi(z-z_0)}+\frac{i\Bn_0}{4\pi(z-\overline{z}_0)}, \eeq{3.2}
will ensure that \eq{1.9} holds and, if  $z_0$ is outside  $C\cup\overline{C}$, $\Real \Bv(t_0)=\Real \Bv_0(t_0)$. Indeed, if $z_0$ is outside  $C\cup\overline{C}$, then, upon application of the residue theorem, the above choice of $\Br(z)$  will lead to the following value of the output function \eq{1.2} at $t=t_0$
\beqa \Real \Bv(t_0)& = & a_0\int_{-1}^1d\BGg(\Gl)\left(\int_{C\cup\overline{C}}\frac{\Br(z)}{\Gl-z}\,dz\right) \nonum
& =& \frac{a_0}{2}\int_{-1}^1d\BGg(\Gl)\left(\frac{\Bn_0}{\Gl-z_0}+\frac{\Bn_0}{\Gl-\overline{z}_0}\right)\nonum
  &= & \Real \Bv_0(t_0),
  \eeqa{3.3}
  and we are finished. } If the curve $C$ traced out by $z(\omega(s))$ is such that $C\cup\overline{C}$ encircles $[-1,1]$ once, clockwise,
  then we need to reverse the sign of $\Br(z)$ in \eq{3.2}. 
  On the other hand, if $z_0$ is inside  $C\cup\overline{C}$, a deformation of this closed curve towards infinity and Cauchy's theorem
  imply the contour integral is zero. Then, $\Real \Bv(t_0)=0$ and we cannot proceed to obtain
  $\Real \Bv_0(t_0)$.
  \section{Measure independent results valid for all time}
  \setcounter{equation}{0}
  \labsect{4}
  In Sections \sect{1} and \sect{3} we obtained results for $\Real v(t_0)$ that only depended on $a_0$ and possibly also the
  first moment of the measure or only on the response at a fixed frequency. Remarkably, and much to our surprise, numerical simulations showed
  that similar results hold for all time $t$ when $\Go(0)$ and $\Go(1)$ both lie on the imaginary axis. 
  To explain this, we take real $a$ and $b$ and assume that
\beq \Go(0) =  -ia,\quad \Go(1)=-ib, \eeq{4.1}
have images
\beq z(\Go(0))=A>1,\quad z(\Go(1))=B<-1, \eeq{4.2}
where \eq{1.5a} implies $A$ and $B$ are real. For $s\in (0,1)$ we assume
$\Go(s)$ traces out a curve $\GG$ in the quadrant $\Real\Go >0$, $ \Imag\Go>0$, while
$z(\Go(s))$ takes values in the upper half plane tracing anticlockwise the curve $C=z(\GG)$ as
$s$ increases from $0$ to $1$.
  
Then, assuming without loss of generality that $t_0=0$, \cb from \eqref{1.2} we have
\beqa \Real \Bv(t)& = & a_0\Real\int_{0}^1\BF_\Gg(z(\omega(s)))\BGa(s)e^{-i\Go(s)t}\,ds \nonum
& = & a_0\Real\int_{0}^1 e^{-i\Go(s)t}\left(\int_{-1}^1\frac{d\BGg(\Gl)}{\Gl-z(\omega(s))}\right)\BGa(s)\,ds. 
\eeqa{4.3}
We choose an applied field with $\BGa(s)$ such that $\Br(z)$, given by \eq{1.8}, takes the form $\Br(z)=-i\Bn_0/(2\pi)$, that is the expression taken by $\Br(z)$ in \eq{1.10a} when $k=0$. \cb Then, switching variables from $s$ to $\Go$, \eq{4.3} implies
\cb
\beq \Real \Bv(t) = a_0\int_{-1}^1d\BGg(\Gl)\left(\Real\frac{\Bn_0}{i\pi}\int_{\GG}e^{-i\Go t}\frac{z'(\omega)}{\Gl-z(\omega)}\,d\omega\right).
\eeq{4.4}
\cb
Additionally substituting $\Gz=i\Go$, $h(\Gz)=z(-i\Gz)$ and letting $D=i\GG$ gives
\cb
\beq \Real \Bv(t)  = a_0\int_{-1}^1d\BGg(\Gl)\left(\Real\frac{\Bn_0}{i\pi}\int_{D}E(\Gz)\frac{h'(\Gz)}{\Gl-h(\Gz)}\,d\Gz\right),
\eeq{4.5}
\cb
where $E(\Gz)=e^{-\Gz t}$. Let us take $\GO$ as the domain inside $D\cup\overline{D}$. Suppose $\Gl=h(\Ga_j)$ with multiplicity $m_j\geq 1$ for $j=1,2,\ldots,m$ with $\Ga_j\in\GO$
and  $h(\Gb_k)=\infty$ with multiplicity $n_k\geq 1$ for $k=1,2,\ldots,n$ with $\Gb_k\in\GO$. Observe that the logarithmic derivative $h'(\Gz)/(h(\Gz)-\Gl)$
has poles at $\Gz=\Gb_k$, $k=1,2,\ldots,n$, with residues $-n_k$, and at $\Gz=\Ga_j$, $j=1,2,\ldots,m$, with residues $m_j$.
Then, noting that $h(\Gz)$ and \cb $E(\Gz)$ \cb are real symmetric 
\beq h(\overline{\Gz})=\overline{h(\Gz)},\quad  E(\overline{\Gz})=\overline{E(\Gz)}, \eeq{4.8}
a simple application of the residue theorem shows that
\beqa \Real\frac{1}{i\pi}\int_{D}E(\Gz)\frac{h'(\Gz)}{\Gl-h(\Gz)}\,d\Gz & = &  \frac{1}{2i\pi}\int_{D\cup\overline{D}}E(\Gz)\frac{h'(\Gz)}{\Gl-h(\Gz)}\,d\Gz \nonum
& = & \sum_{k=1}^{n}  n_k E(\Gb_k)-\sum_{j=1}^{m(\Gl)} m_j(\Gl)E(\Ga_j(\Gl)), \nonum &~&
\eeqa{4.9}
where  we replaced $m$, $m_j$ and $\Ga_k$ with $m(\Gl)$,  $m_j(\Gl)$ and $\Ga_k(\Gl)$ to emphasize their dependence on $\Gl$.
Note that \eq{4.8} implies that the points $\Ga_j$ and $\Gb_k$ come in complex conjugate pairs and this ensures that the right hand
side of \eq{4.9} is real. This relation \eq{4.9} holds for any analytic function \cb $E(\Gz)$ \cb defined in an open neighborhood of the closure of $\GO$
that is real symmetric, and not just $E(\Gz)=e^{-\Gz t}$.  
Thus,
\cb
\beq \Real \Bv(t)=a_0\Bn_0\sum_{k=1}^{n} n_k e^{-t\Gb_k}-a_0\int_{-1}^1d\BGg(\Gl)\Bn_0\sum_{j=1}^{m(\Gl)} m_j(\Gl)e^{-t\Ga_j(\Gl)}
\eeq{4.12}
\cb
will be measure independent for all time $t$ if and only if $h(\Gz)=z(-i\Gz)$ does not take real values in
$[-1, 1]$ throughout $\GO$. If it is measure independent, then we can obtain the volume fraction $f_1=a_0/2$ from measurements
of $\Real \Bv(t)$ at any time $t$.

If  $m(\Gl)$ and  $m_j(\Gl)$ do not depend on $\Gl$, then at $t=t_0=0$ \eq{4.9} implies \cb
\beq \frac{1}{2\pi}\Real\int_{D\cup\overline{D}}\frac{h'(\Gz)}{\Gl-h(\Gz)}\,d\Gz
= \sum_{k=1}^{n} n_k -\sum_{j=1}^{m}m_j,
\eeq{4.12a}
\cb and the right hand side can be identified with the winding number of $C\cup\overline{C}=h(D\cup\overline{D})$
about $z=\Gl$. This is $\pm 1$ if $C$ lies in the upper half plane and  $C\cup\overline{C}$ encloses the
interval $[-1,1]$ and we get agreement with \eq{1.11}. We assumed $D$ is traced anticlockwise around $(a+b)/2$
as $s$ is increased from $0$ to $1$. This will not be the case if the right hand side of \eq{4.12a} is $-1$. Rather
$D$ will be traced clockwise so that $C$ is traced anticlockwise around the origin. This will reverse the sign of \eq{4.12a}
giving agreement with \eq{1.11}.

Note that if $\Gm_2$ (or $\Gm_1$) is real, positive, and frequency independent, then as observed in Section \sect{2}, \eq{2.3} implies $z(\Go)$ will have
strictly positive (or negative, respectively) imaginary part when $\Go$ is in the quadrant $\Real \Go>0, \Imag \Go>0$. Thus, the poles $\Gb_k $ of $h(\Gz)$ and zeros
$\Ga_j$ of $h(\Gz)-\Gl$
will be simple and alternate along the real axis.

One can also get time independent results that only incorporate the first moment of the measure. \cb Again, \cb we take an applied field with $\BGa(s)$ such that
$\Br(z)=-i\Bn_0z/(2\pi)$ \cb but this time we set $E(\Gz)=h(\Gz)e^{-\Gz t}$\cb. Now $E(\Gz)$ is not analytic in an open neighborhood of the closure of $\GO$,
but has poles at the points  $\Gb_k, k=1,2,\ldots,n$. However, we have
\beq  \frac{h'(\Gz)h(\Gz)}{\Gl-h(\Gz)}=-h'(\Gz)  + \frac{\lambda h'(\Gz)}{\lambda-h(\Gz)}. \eeq{4.22a}
If we assume simple poles, i.e. $n_k=1$ for all k, then near the pole at $\Gb_k$, $h(\Gz)$ has an expansion
\beq h(\Gz)=\frac{b_k}{\Gz-\Gb_k}+c_k+d_k(\Gz-\Gb_k)+\ldots, \eeq{4.23}
giving
\beq \frac{h'(\Gz)h(\Gz)e^{-\Gz t}}{\Gl-h(\Gz)}=\frac{b_k e^{-\Gb_k t}}{(\Gz-\Gb_k)^2}-\frac{b_k t e^{-\Gb_k t}}{\Gz-\Gb_k}+
\frac{\Gl e^{-\Gb_k t}}{\Gz-\Gb_k}+\ldots. \eeq{4.24}
Consequently, we get
\cb
\beqa \Real \Bv(t)& = &
a_0\sum_{k=1}^{n} e^{-t\Gb_k}\int_{-1}^1\Gl\,d\BGg(\Gl)\Bn_0- a_0\Bn_0\sum_{k=1}^{n}b_k t e^{-\Gb_k t}\nonum
&~&-a_0\int_{-1}^1\sum_{j=1}^{m(\Gl)} m_j(\Gl)\Gl e^{-t\Ga_j(\Gl)}\,d\BGg(\Gl)\Bn_0.
\eeqa{4.25}
\cb
This will be measure dependent for all times $t\ne t_0$ if $z(-i\Gz)$ does take real values in
$[-1, 1]$ throughout $\GO$. Otherwise, the result depends for all time only on the measure through its first moment.

\cb              
\section{Revisiting the problem of recovering the response at any
frequency} \cb
\setcounter{equation}{0}
\labsect{3a}

Now suppose we choose an applied field with $\BGa(s)$ such that $\Br(z)$ is given by \eq{3.2},
in which $z_0=z(\Go_0)$ where $\Go_0$ is some, possibly complex, frequency at which we desire to determine the response. We assume $z_0$ does not
lie on the interval $[-1,1]$ on the real axis, where $\Gl$ lies.
We let
\beq E(\Gz)=\frac{e^{-\Gz t}}{2(h(\Gz)-z_0)}+\frac{e^{-\Gz t}}{2(h(\Gz)-\overline{z_0})}.
\eeq{4.26}
This will have poles at points $z_0=h(\Gk_\ell)$ with multiplicity $p_\ell\geq 1$ for $\ell=1,2,\ldots,p$ with $\Gk_\ell\in\GO$,
and poles at points $\overline{z_0}=h(\overline{\Gk_\ell})$ with multiplicity $p_\ell\geq 1$ for $\ell=1,2,\ldots,p$ with $\overline{\Gk_\ell}\in\GO$.
When $z_0$ is real, the $\Gk_\ell$ include their complex conjugates. Observing that
\beqa &~&E(\Gz)\frac{h'(\Gz)}{\Gl-h(\Gz)} =
\left[\frac{1}{2(\Gl-z_0)}+\frac{1}{2(\Gl-\overline{z_0})}\right]\left[\frac{h'(\Gz)e^{-\Gz t}}{\Gl-h(\Gz)}\right]\nonum
&~&\quad  +\frac{1}{2(\Gl-z_0)}\left[\frac{h'(\Gz)e^{-\Gz t}}{h(\Gz)-z_0}\right]
  +\frac{1}{2(\Gl-\overline{z_0})}\left[\frac{h'(\Gz)e^{-\Gz t}}{h(\Gz)-\overline{z_0}}\right],
\eeqa{4.27}
and proceeding as before, we obtain for complex $z_0$ that
\beqa
&~& \Real\frac{1}{i\pi}\int_{D}E(\Gz)\frac{h'(\Gz)}{\Gl-h(\Gz)}\,d\Gz \nonum
&~&=\quad -\left[\sum_{j=1}^{m(\Gl)}m_j(\Gl)e^{-t\Ga_j(\Gl)}\right]\left[\frac{1}{2(\Gl-z_0)}+\frac{1}{2(\Gl-\overline{z_0})}\right]
\nonum
&~&\quad +\left[\sum_{\ell=1}^{p} p_\ell e^{-t\Gk_\ell}\right]\frac{1}{2(\Gl-z_0)}+\left[\sum_{\ell=1}^{p} p_\ell e^{-t\overline{\Gk_\ell}}\right]\frac{1}{2(\Gl-\overline{z_0})},
\eeqa{4.28}
\cb where we have assumed the integral over $D$ is anticlockwise. This gives 
\beqa
\Real\Bv(t)& = &
a_0\int_{-1}^1d\BGg(\Gl)\Bn_0\sum_{j=1}^{m(\Gl)}m_j(\Gl)e^{-t\Ga_j(\Gl)}\left[\frac{1}{2(\Gl-z_0)}+\frac{1}{2(\Gl-\overline{z_0})}\right] \nonum
&~&-a_0\sum_{\ell=1}^{p} p_\ell e^{-t\Gk_\ell}\int_{-1}^1\frac{d\BGg(\Gl)\Bn_0}{2(\Gl-z_0)} \nonum
&~&-a_0\sum_{\ell=1}^{p} p_\ell e^{-t\overline{\Gk_\ell}}\int_{-1}^1\frac{d\BGg(\Gl)\Bn_0}{2(\Gl-\overline{z_0})}.
\eeqa{4.29}
If $z(-i\Gz)$ does not take real values in $[-1, 1]$ throughout $\GO$ ($m=0$) and $p\ne 0$ then we can recover the response at frequency $\Go_0$ from measurements at any two times $t_1$ and $t_2\ne t_1$. To see this, we take $p=1$, which as we will
see shortly is necessarily the case when $m=0$ and $p\ne 0$, and set
\beq \Gx=\int_{-1}^1\frac{d\BGg(\Gl)\Bn_0}{2(\Gl-z_0)}. \eeq{4.29.1}
Then \eq{4.29} implies the linear equations 
\beqa \Real\Bv(t_1)& = & -a_0\Gx e^{-t_1\Gk_1}-a_0\overline{\Gx} e^{-t_1\overline{\Gk_1}} \nonum
\Real\Bv(t_2)& = & -a_0\Gx e^{-t_2\Gk_1}-a_0\overline{\Gx} e^{-t_2\overline{\Gk_1}},
\eeqa{4.19.2}
which can be solved for $\Gx$, and hence its real part, giving $v(t_0)$. On the other hand, if $z(-i\Gz)$ takes real values in $[-1, 1]$ inside $\GO$, then
we cannot recover the response at frequency $\Go_0$, for all measures $\BGg$, unless $t=0$.  

When $z_0$ is real, \eq{4.29} should be replaced by
\beqa
\Real\Bv(t)& = & a_0\int_{-1}^1d\BGg(\Gl)\Bn_0\sum_{j=1}^{m(\Gl)}m_j(\Gl)e^{-t\Ga_j(\Gl)}\left[\frac{1}{\Gl-z_0}\right] \nonum
&~&-a_0\left[\sum_{\ell=1}^{p} p_\ell e^{-t\Gk_\ell}\right]\int_{-1}^1\frac{d\BGg(\Gl)\Bn_0}{\Gl-z_0},
\eeqa{4.30}
and for $t\ne 0$ and $p\ne 0$ we can recover the response at frequency $\Go_0$ for all measures $\BGg$ if and only if $z(-i\Gz)$ does not take real values in
$[-1, 1]$ throughout $\GO$ ($m=0$).

Since the point $z_0$ lies outside $C\cup\overline{C}$, the winding number of $C\cup\overline{C}$ around $z_0$ is zero, that is:
\beq  \sum_{\ell=1}^{p} p_\ell -\sum_{k=1}^n n_k=0. \eeq{7.1}
On the other hand, assuming the function $h:D\to C$ is orientation preserving,
the winding number of  $C\cup\overline{C}$ around the point $\lambda$ is equal to one:
\beq \sum_{j=1}^{m(\lambda)} m_j(\lambda)-\sum_{k=1}^n n_k=1. \eeq{7.2}
This implies $m\ne 0$, and we cannot recover $\Real\Bv(t_0)$ from measurements of $\Real\Bv(t)$ except at $t=0$.
However, if the mapping $h:D\to C$ reverses orientation, then
\beq \sum_{j=1}^{m(\lambda)} m_j(\lambda)-\sum_{k=1}^n n_k=-1, \eeq{7.2a}
and $m$ can only be zero if $n=p=1$. 

By setting $t=0$ in \eq{4.29} and \eq{4.30} and using these identities we recover the conclusion (\ref{3.3})
of section \sect{3}. Note that if $h:D\to C$ reverses orientation, we need to change signs of the right hand sides of \eq{4.29} and \eq{4.30}.
If $C$ reverses orientation this sign change is a consequence of the sign change of $r(z)$ given by \eq{3.2} needed to ensure \eq{3.3} holds.
Alternatively if $D$ reverses orientation, then the right hand side of \eq{4.28} changes sign and there is no change in the sign of $r(z)$. \cb

Throughout all cases discussed above it was implicitly assumed that the curve $D$ is simple, that is, without self-intersections. \cb Therefore, \cb $D \cup \overline{D}$ is a Jordan curve, the boundary 
of the domain $\Omega$. We can relax this condition, simply assuming that $D$ is a curve contained in the upper half-plane, with boundaries at the real points.
Leaving aside the domain $\Omega$, we still can invoke Cauchy's theorem in this more general setting, with the necessary modification of the weights $n_k, m_j(\lambda), p_\ell$. Specifically, these numbers will consist of the respective multiplicities, times the winding number with respect to $D \cup \overline{D}$ of the associated zero or pole. Then it may well happen that some winding numbers are equal to zero, thus erasing the contribution of the respective singularity of the integrand.

\cb
\section{Numerical results}
\setcounter{equation}{0}

Let us start by considering a composite made of two phases, one of which, say phase 2, is lossless, so that $z(\omega)$ takes the form \eqref{2.3}. Furthermore, assume that $\mu_1(\omega)=1+i/\omega$, thus mimicking the low frequency dielectric response of a lossy
dielectric material, and $\mu_2=2$. According to the procedure illustrated in Section \sect{2}, we can generate an appropriate trajectory $\Gamma$ traced by $\omega(s)$, with $s\in[0,1]$, by taking two frequencies $\omega_1$ and $\omega_2$ on opposite sides of the frequency $\omega_*$ for which $\mu_1(\omega_*)=\mu_2$. In this example, $\omega_*=i$, so let us take $\omega_1=1.5i$ and $\omega_2=0.5i$. We verify  that $z(\omega_2)=-5<-1$ and $z(\omega_1)=11>1$, so, then, we can take a trajectory $\Gamma$ linking $\omega_1$ and $\omega_2$ that remains in the quadrant $\Real \omega>0$, $\Imag \omega>0$. Consider, for instance, $\omega(s)=-2(1+i)s^2+(2+i)s+1.5i$, for which $\omega(0)=\omega_1$ and $\omega(1)=\omega_2$. Then, $z(\omega(s))$ takes values in the upper half plane tracing anticlockwise the curve $C=z(\Gamma)$ as $s$ increases from 0 to 1, so that the curve $C\cup\overline{C}$ is a closed curve encircling the interval $[-1,1]$ once, see \protect{\fig{fig:curves_model1}(a)}.

\begin{figure}[h!]
	\begin{subfigure}[t]{0.5\textwidth}
		\centering
		\includegraphics[width=\linewidth]{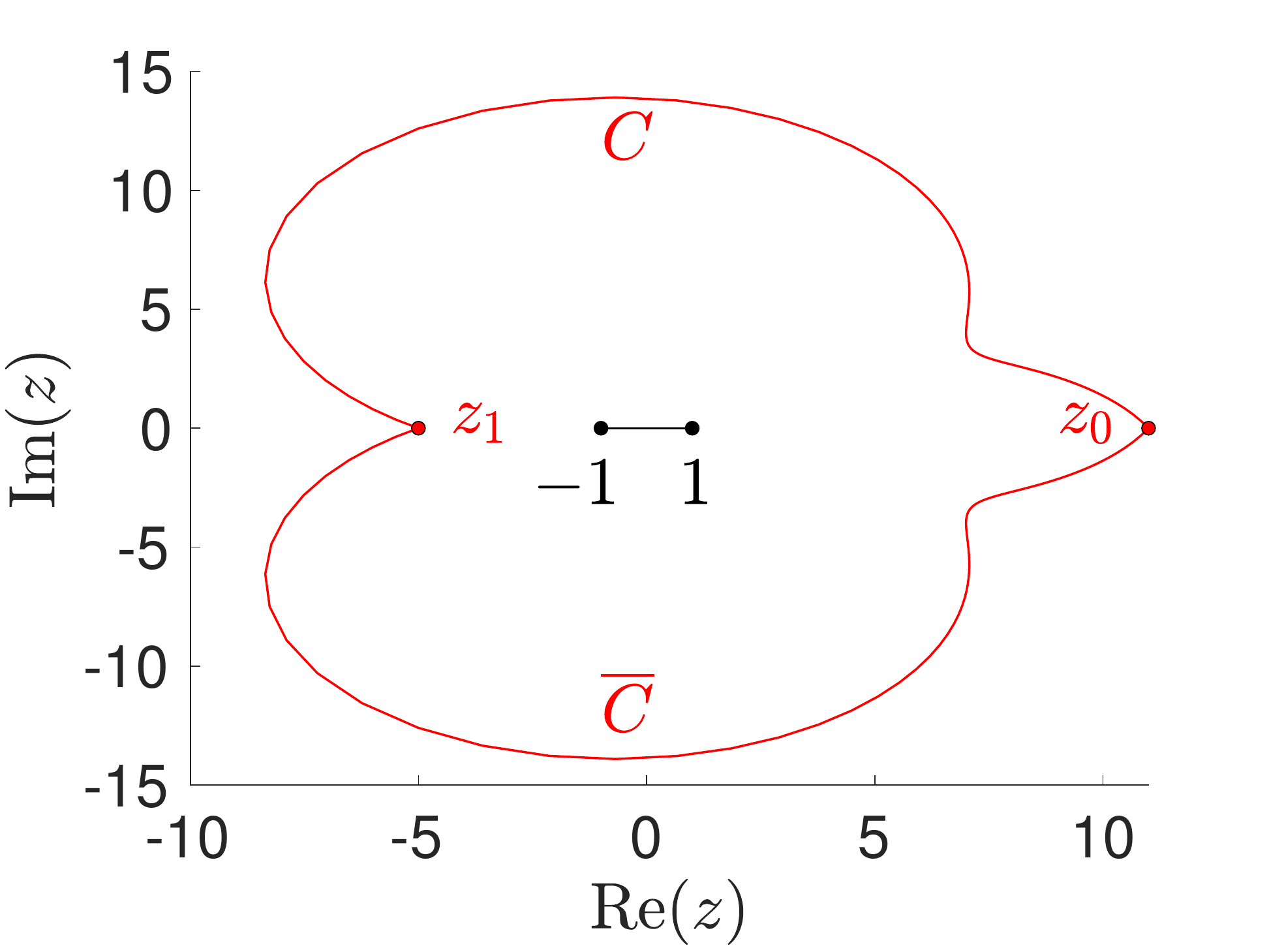}
		\caption{}
	\end{subfigure}	
	\begin{subfigure}[t]{0.5\textwidth}
		\centering
		\includegraphics[width=\linewidth]{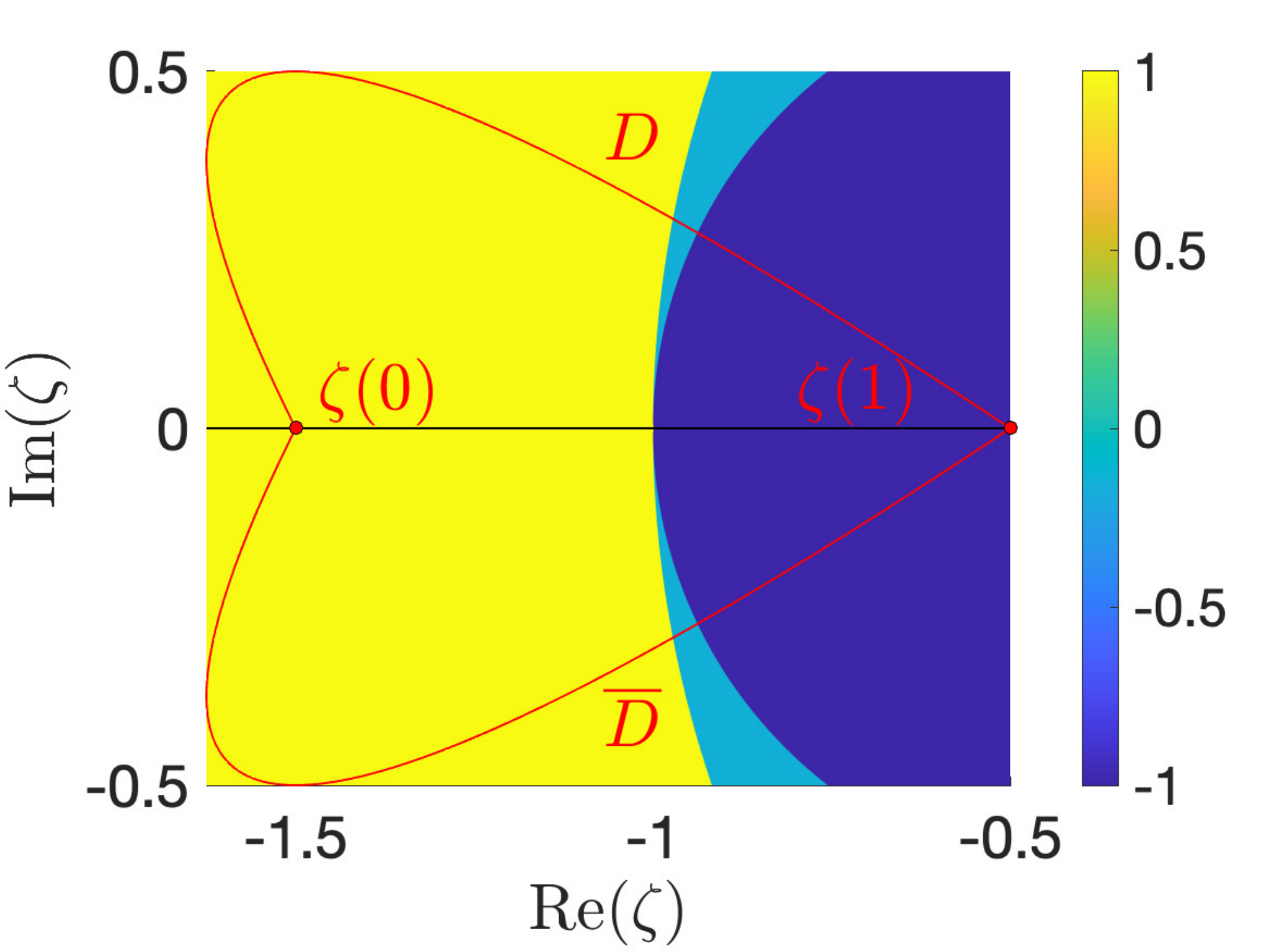}
		\caption{}
	\end{subfigure}
	\caption{(a) Given $\mu_1=1+i/\omega$ and $\mu_2=2$, we choose the trajectory $\omega(s)=-2(1+i)s^2+(2+i)s+1.5i$ with $s\in[0,1]$, so that its image through the function $z(\omega)$ given by \eqref{2.3} is the curve connecting the points $z_0=z(\omega(0))=11$ and $z_1=z(\omega(1))=-5$. Then, the curve $C\cup\overline{C}$ is a closed curve encircling the interval $[-1,1]$ once. (b) In red is the curve $D$ traced by  $\Gz(s)=i\omega(s)$ as $s$ is increased from 0 to 1, and the curve $\overline{D}$ traced by $\overline{\Gz(s)}$. The domain inside $D\cup\overline{D}$ is $\Omega$. In black is represented the line where the function  $h(\Gz)=z(-i\Gz)$  takes real values, the yellow region is where the real part of $h(\Gz)$ takes values bigger than 1 and  the purple region where it takes values smaller than -1. The color bar indicates that the real part of $h(\Gz)$ never takes values $[-1,1]$ in the domain $\Omega$, especially along the black line, where the imaginary part of  $h(\Gz)$ is zero. Notice that the function has a single pole at $\zeta=-1$: indeed, $z(\omega)$ has a pole at $\omega_*=i$.}
	\labfig{fig:curves_model1}
\end{figure}

Note that the trajectory chosen is rather special \cb because the function $h(\Gz)=z(-i\Gz)$ does not take real values in $[-1,1]$ throughout $\Omega$ (the domain inside $D\cup\overline{D}$), see \protect{\fig{fig:curves_model1}(b)}. This ensures that the bounds on $\Real{\bf v}(t)$ are measure-independent not only at time $t=t_0$, but  for any time $t$. \cb
To see this, for simplicity, we consider the one-dimensional forward problem in which the input function $\Real{\bf u}(t)$ \eqref{1.1} has only one non-zero component, that we will denote by $\Real{u}(t)$, and we look at the response \eqref{1.2} of the material in the same direction, that we will denote with $\Real{v}(t)$. First, we assume that the first moment of the measure is not known, and set $k=0$ in the formula \eqref{1.10a} that determines  the relevant component of the function $\Br$. The corresponding input function is depicted in \protect{\fig{fig:bounds_Re_v}(a)}, whereas \protect{\fig{fig:bounds_Re_v}(b)} shows the bounds on the output function $\Real{v}(t)$.

For any moment of time, the bounds are found  by optimizing over the measure: the maximum and minimum values are attained when the relevant component of the matrix-valued measure  $\boldsymbol{\gamma}(\lambda)$, that we denote with ${\gamma}(\lambda)$, is an extreme measure, namely the point mass ${\gamma}(\lambda)=\delta(\lambda-\lambda_0)$, where $\lambda_0$ is varied over $[-1,1]$. The outer bounds, in blue, in  \protect{\fig{fig:bounds_Re_v}(b)} correspond to the case when the volume fraction is not given (note that the upper bound is always zero), and therefore $a_0$, is not known, whereas the inner ones do incorporate the volume fraction ($a_0=0.6$). Notice that the latter are indeed coincident and not only the value they take at $t=t_0=0$ equals $a_0$, as predicted by \eqref{1.11}, but the value they take at any moment of time $t$ is exactly the one provided by \eqref{4.12}, where $\beta_1=-1$ is the only simple pole ($n_1=1$) of the function $h(\zeta)$ in $\Omega$, and the second part of the formula is zero as $h(\zeta)$ does not have any zero in the domain $\Omega$, see also \protect{\fig{fig:curves_model1}(b)}. Therefore, according to \eqref{4.12}, the two coincident bounds incorporating the volume fraction have the following analytical expression: \(\Real v(t)=-a_0\exp(t)\), which is indeed in agreement with the result \cb shown \cb in \protect{\fig{fig:bounds_Re_v}(b)}. Note that any other trajectory $\omega(s)$ having the same features of the one chosen in this example, that are, the corresponding \cb closed curve \cb $C\cup\overline{C}$ encircles $[-1,1]$, and the corresponding \cb closed curve \cb  $D\cup\overline{D}$ is such that \cb $h(\zeta)$ \cb does not take real values in $[-1,1]$ inside it, would have led to the same bounds depicted in \protect{\fig{fig:bounds_Re_v}(b)}.
\begin{figure}[h!t]
	\begin{subfigure}[t]{0.5\textwidth}
		\centering
		\includegraphics[width=\linewidth]{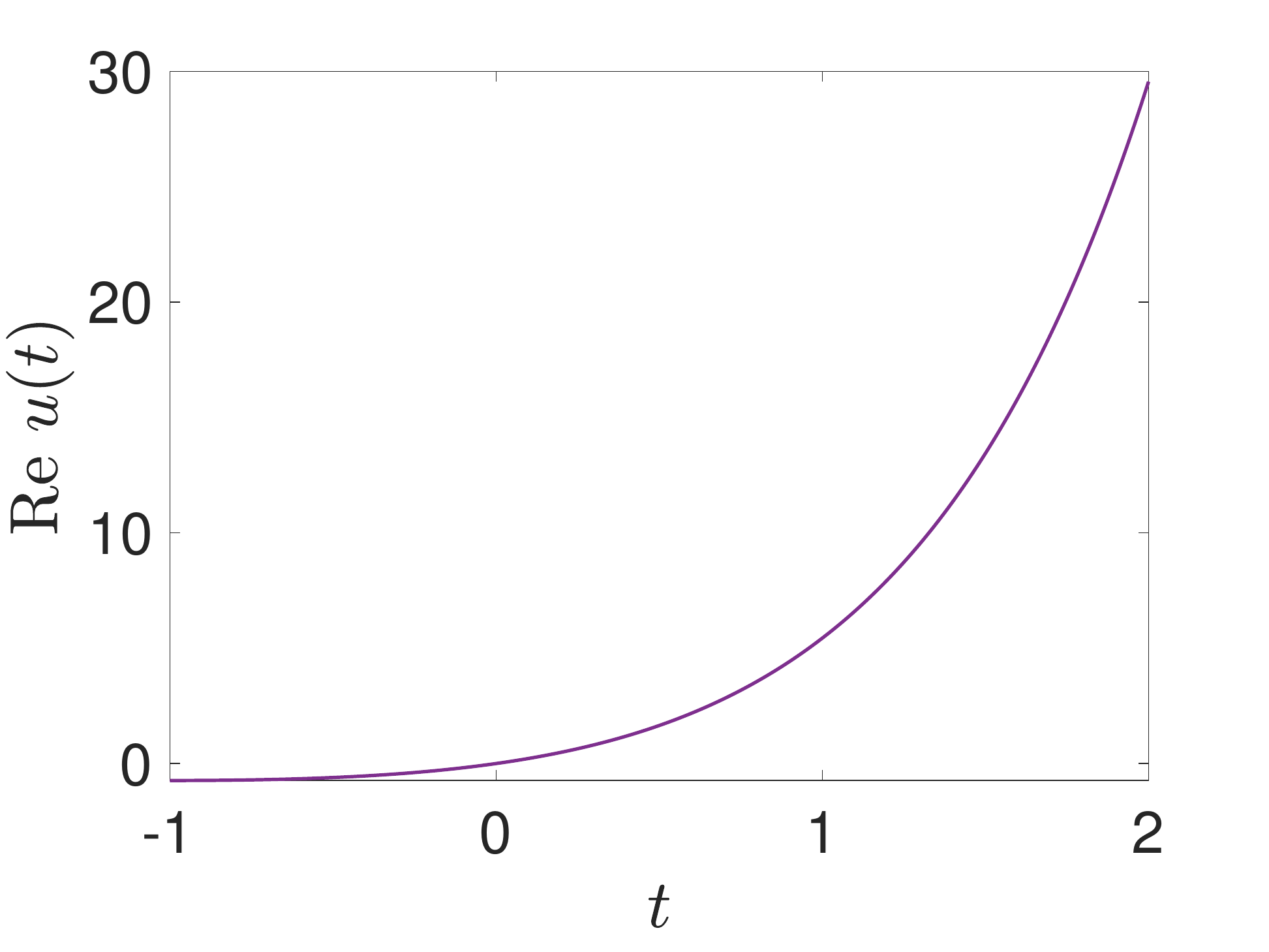}
		\caption{}
	\end{subfigure}	
	\begin{subfigure}[t]{0.5\textwidth}
		\centering
		\includegraphics[width=\linewidth]{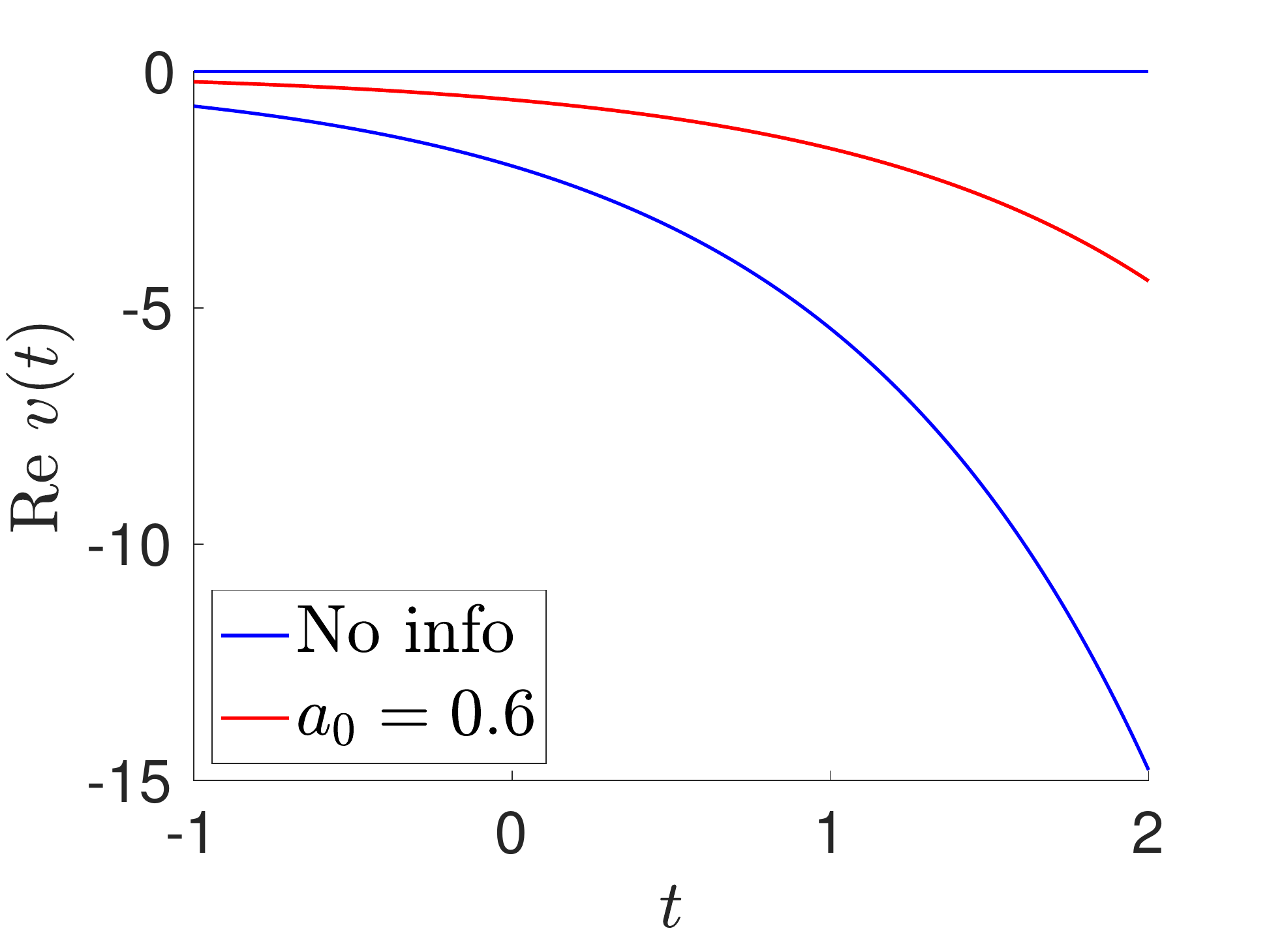}
		\caption{}
	\end{subfigure}
	\caption{We take $\mu_1=1+i/\omega$, $\mu_2=2$, and $\omega(s)=-2(1+i)s^2+(2+i)s+1.5i$. (a) The relevant component of the input function \eqref{1.1}, $\Real[u(t)]$, is found by choosing the relevant component of the function $\Br$ according to \eqref{1.10a}, with $k=0$. Therefore, $\alpha(s)$ is given by \eqref{1.8}, and then the coefficient $\beta(s)$ appearing in \eqref{1.1} is known through \eqref{1.5}. (b) By optimizing over the measure with one point mass for any moment of time, we obtain bounds on the output function $\Real[v(t)]$ in two different scenarios: in blue are the bounds when the volume fraction, and therefore $a_0$, is not given, whereas in red are the bounds incorporating the volume fraction $f_1$ ($a_0=2f_1=0.6$). The upper blue bound is clearly always zero, whereas the lower blue bound takes value -2 at $t=0$ and decreases in time. Interestingly enough, the red bounds are coincident: indeed,
          the spectrum of frequencies $\omega(s)$ chosen is such that the bounds are measure-independent at any moment of time. Furthermore, their analytical expression is exactly determined by equation \eqref{4.12} which, upon substituting the value $\beta_1=-1$ of the only simple pole, $n_1=1$,  of the function $h(\zeta)$ in $\Omega$,  provides $\Real[v(t)]=-a_0\exp(t)$ (according to \protect{\fig{fig:curves_model1}(b)}, $h(\zeta)$ does not have any zero in $\Omega$ and the curve $D$ is traced clockwise).}
	\labfig{fig:bounds_Re_v}
\end{figure}

Now suppose that we know the first moment of the measure: $M_1=\int_{-1}^1\lambda d{\gamma}(\lambda)$. Then, we apply an input function $\Real{u}(t)$ so that the corresponding output function will incorporate such a piece of information by choosing, for instance, $k=1$ in \eqref{1.10a}. Therefore, $\alpha(s)$ \eqref{1.8} and $\beta(s)$ \eqref{1.5} are known and the input function \eqref{1.1} is determined, see \protect{\fig{fig:bounds_Re_v_moments}(a)}. We compute the bounds on the output function $\Real{v}(t)$, by considering the extreme measure $\gamma(\lambda)=w_0\delta(\lambda-\lambda_0)+w_1\delta(\lambda-\lambda_1)$, where the weights $w_0$ and $w_1$ are chosen so that the values of the zeroth and first moments of the measure are the desired ones, and $\lambda_0$ and $\lambda_1$  are varied over the interval $[-1,1]$ in search for the minimum and maximum values of  $\Real{v}(t)$ at any time $t$, see  \protect{\fig{fig:bounds_Re_v_moments}(b)}. The outer bounds in blue correspond to the case where only the volume fraction is known, whereas the inner bounds in red incorporate also the first moment of the measure. Like in the previous case, the latter are coincident, due to the special choice of the trajectory $\omega(s)$, and their analytical expression is given by the sum of the relations \eqref{4.12} (which does not include $M_1$) and \eqref{4.25} (which includes $M_1$ only). For the case under study, $h(\zeta)$ has only one single pole, $\beta_1=-1$ with corresponding residue $b_1=-4$, and no zeros in the domain $\Omega$. Therefore, the analytical expression of the output function, when $M_1=0.4$, is \(\Real v(t)=-a_0\exp(t)(1.4+4t)\) which is, indeed, in agreement with the numerical results in \protect{\fig{fig:bounds_Re_v_moments}(b)}.
\begin{figure}[h!t]
	\begin{subfigure}[t]{0.5\textwidth}
		\centering
		\includegraphics[width=\linewidth]{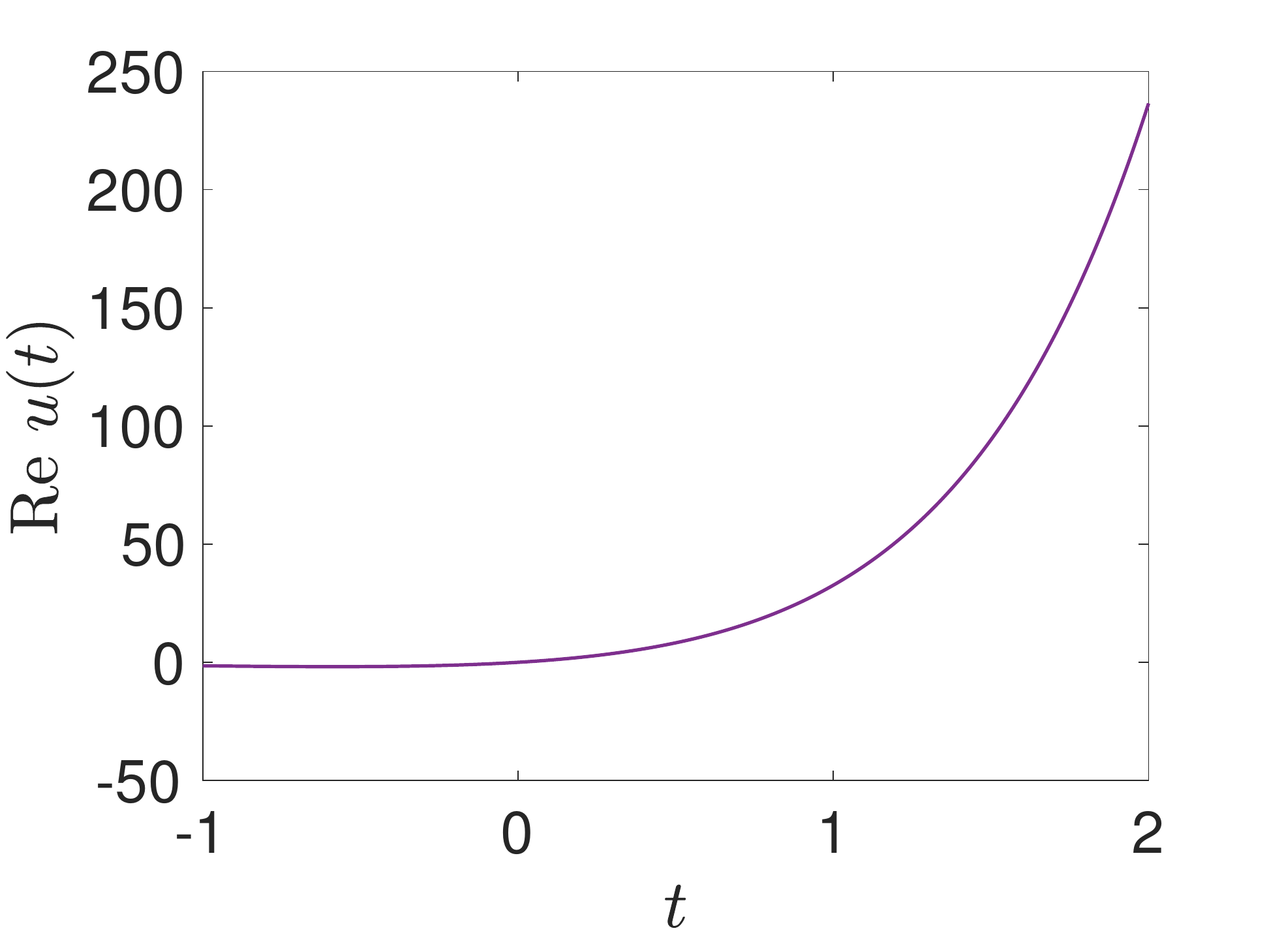}
		\caption{}
	\end{subfigure}	
	\begin{subfigure}[t]{0.5\textwidth}
		\centering
		\includegraphics[width=\linewidth]{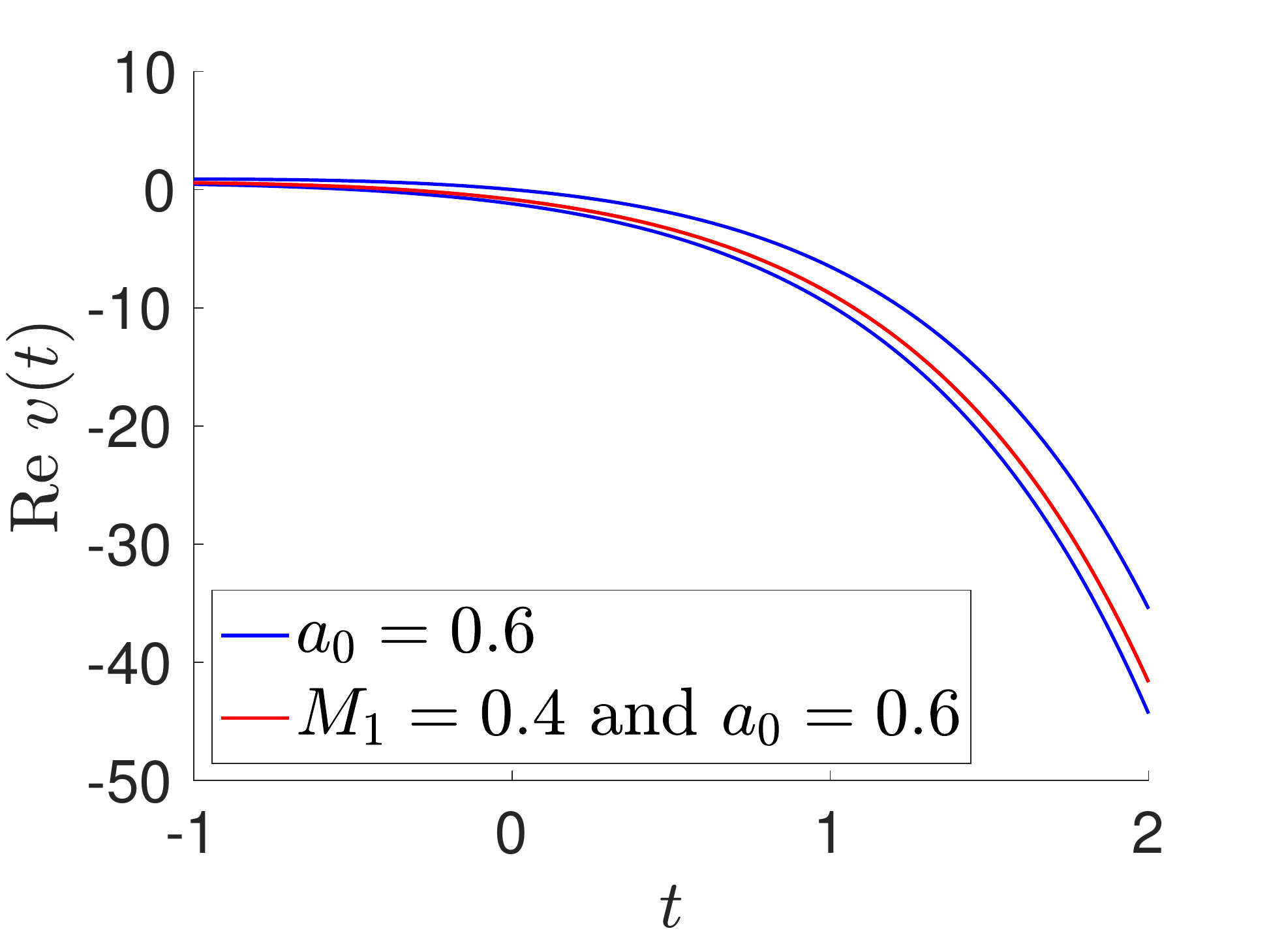}
		\caption{}
	\end{subfigure}
	\caption{We take $\mu_1=1+i/\omega$, $\mu_2=2$, and $\omega(s)=-2(1+i)s^2+(2+i)s+1.5i$. Furthermore, we assume that the first moment of the measure $M_1$ is known, say $M_1=0.4$. (a) The input function $\Real[u(t)]$ is found by setting $k=1$ in \eqref{1.10a}. Therefore, $\alpha(s)$ is given by \eqref{1.8}, and then the coefficient $\beta(s)$ appearing in \eqref{1.1} is known through \eqref{1.5}. (b) \cb In blue are the bounds \cb when only $a_0=2f_1=0.6$ is given, whereas in red are the bounds incorporating also the first moment of the measure $M_1=0.4$. The upper blue bound is clearly always zero, whereas the lower blue bound takes value $-a_0$ at $t=t_0=0$ and decreases in time. Again, the red bounds incorporating the first moment of the measure and the volume fraction are coincident: like in the case when the first moment of the measure was not known, the spectrum of frequencies $\omega(s)$ chosen is such that the bounds are independent of the measure, \cb aside from its first moment, \cb at any moment of time.}
	\labfig{fig:bounds_Re_v_moments}
\end{figure}

Finally, suppose that we want to determine the response of the material $\Real v_0(t)$ at a very specific frequency, $\omega_0$, by applying the spectrum of frequencies $\omega(s)$ chosen earlier. As explained in Section \sect{3}, this is only possible if $z(\omega_0)$ is outside $C\cup\overline{C}$. This is true if, for instance, we choose $\omega_0=29/27i$, for which $z(\omega_0)=30$, see \protect{\fig{fig:curves_model1}(a)}. Then, by choosing the input function given by the choice \eqref{3.2}, the corresponding output function  \cb $\Real v(t)$ \cb is such that at $t=t_0$ it provides the value $\Real v_0(t_0)$, as \cb shown \cb by \eqref{3.3}. Such a result holds true for any measure $\gamma$. In  \protect{\fig{fig:bounds_Re_v_momentsN}(a)}, we plot the bounds on both output functions, $\Real v_0(t)$  and $\Real v(t)$, by optimizing over the position of the point mass $\lambda_0$ in $\gamma=\delta(\lambda-\lambda_0)$. Not only the bounds are coincident at $t=t_0=0$, in agreement with \eqref{3.3}, but they are coincident for any moment of time. Notice that, if the measure is known, then the two output functions take exactly the same value at any moment of time $t$, as \cb shown \cb in \protect{\fig{fig:bounds_Re_v_momentsN}(b)}.  Since $z_0$ is real, the analytical expression of the output function is given by \eqref{4.30}, where the first term equals zero, given that $h(\zeta)$ does not take any real value between $-1$ and $1$ in $\Omega$, see \protect{\fig{fig:curves_model1}(b)}, and the second term reads $-a_0\exp(31/27t)/(\lambda_0-z(\omega_0))$.
\begin{figure}[h!t]
	\begin{subfigure}[t]{0.5\textwidth}
		\centering
		\includegraphics[width=\textwidth]{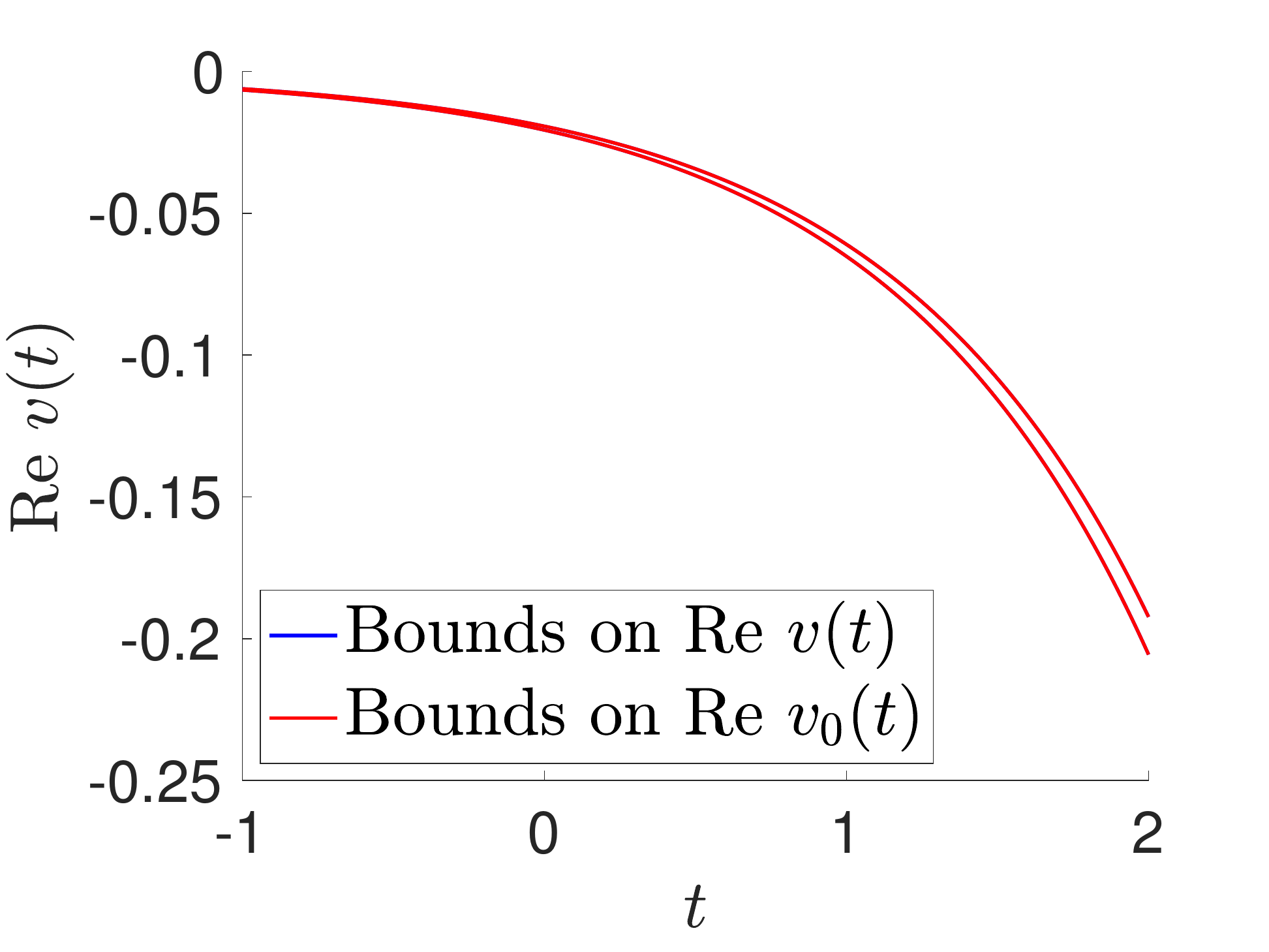}
		\caption{}
	\end{subfigure}	
	\begin{subfigure}[t]{0.5\textwidth}
		\centering
		\includegraphics[width=\linewidth]{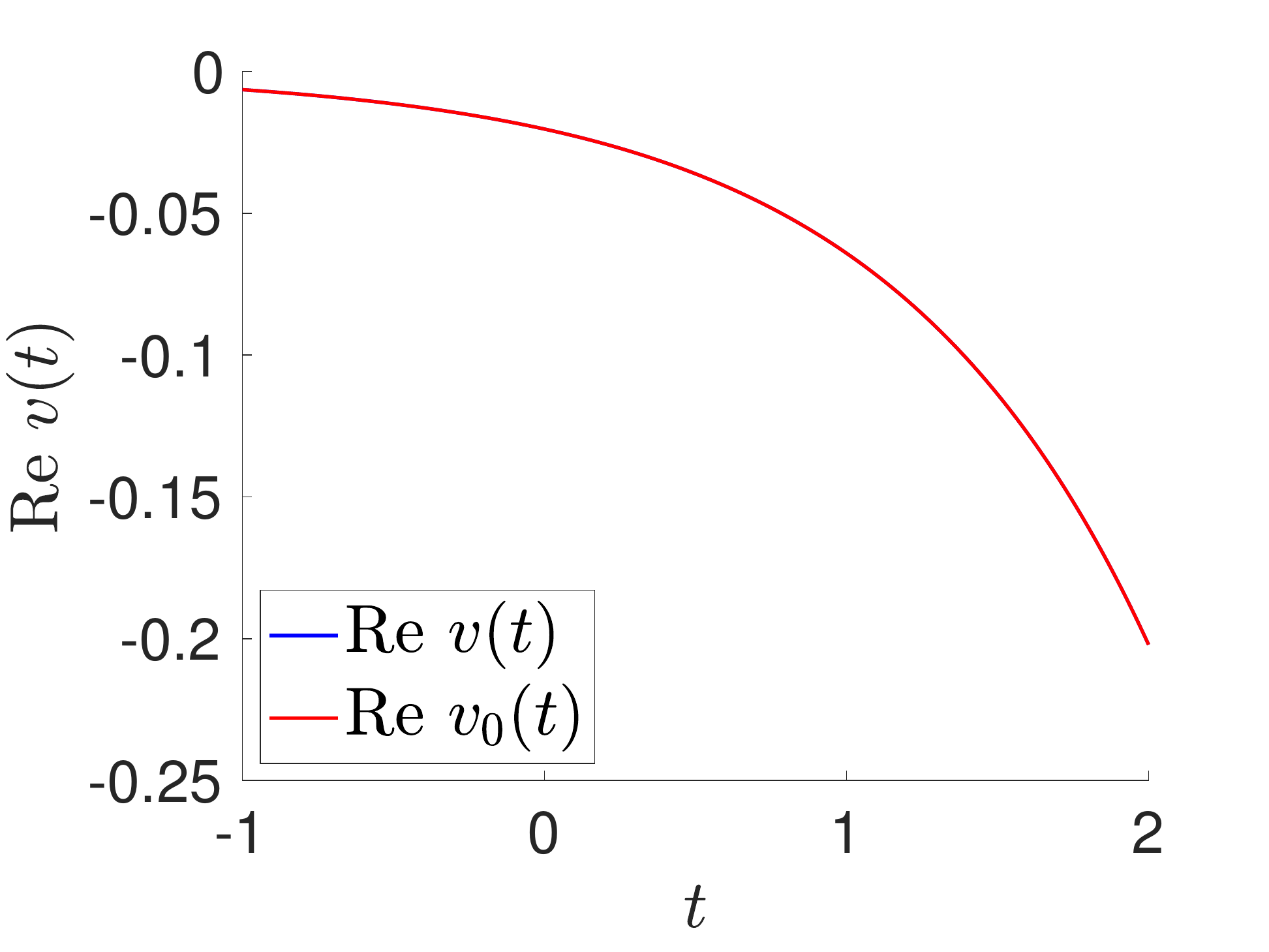}
		\caption{}
	\end{subfigure}
	\caption{We take $\mu_1=1+i/\omega$, and $\mu_2=2$. The value at $t=t_0$ of the output function $\Real[v_0(t)]$, corresponding to an input function at the frequency $\omega_0=29/27 i$, is determined by applying an input function of the type \eqref{1.1} with $\omega(s)=-2(1+i)s^2+(2+i)s+1.5i$. We choose  $\beta(s)$  in \eqref{1.1} by selecting  $\alpha(s)$ in \eqref{1.5} according to \eqref{1.8}, with $r(z)$ given by \eqref{3.2}. (a) Here the measure is unknown and bounds on $\Real[v_0(t)]$, in red,  and $\Real[v(t)]$, in blue, are computed by optimizing over all measures. The bounds coincide not only at $t=t_0=0$ but any moment of time $t$. (b) Here the measure is known to be \cb $\gamma(\lambda)=\delta(\lambda-0.5)$: the value at any $t$ of $\Real[v_0(t)]$, corresponding to an input at the frequency $\omega_0$,  is equal to the value of  $\Real[v(t)]$, corresponding to the input \eqref{1.1} with $\omega(s)$ chosen as described above. The analytical expression is $-a_0\exp(31/27t)/(\lambda_0-z(\omega_0))$.}
	\labfig{fig:bounds_Re_v_momentsN}
\end{figure}

Similar results hold when we consider different models for  the material properties $\mu_1$ and $\mu_2$. \cb Suppose, for instance, that both materials are dispersive,
meaning that they both are frequency dependent. \cb \cb Assume that $\mu_1(\omega)=1-1/\omega^2$, thus mimicking the response of plasma in the dielectric problem, and $\mu_2(\omega)=1+i/\omega$. An appropriate trajectory $\omega(s)$ that generates a curve $C$ so that $C\cup\overline{C}$ is a closed curve that encircles the interval $[-1,1]$ once is the one chosen in the previous example, that is $\omega(s)=-2(1+i)s^2+(2+i)s+1.5i$, as \cb shown \cb in \protect{\fig{fig:curves_model2}(a)}. Furthermore, the chosen trajectory is such that  the curve $D$ traced clockwise by  $\Gz(s)=i\omega(s)$ is such that the function $h(\Gz)=z(-i\Gz)$ does not take real values in $[-1,1]$ throughout $\Omega$ (the domain inside $D\cup\overline{D}$), as \cb shown \cb in \protect{\fig{fig:curves_model2}(b)}. This ensures that the bounds on $\Real{\bf v}(t)$ when the volume fraction is known are measure-independent not only at time $t=t_0$, but  for any time $t$, as \cb shown \cb in \protect{\fig{fig:curves_model2}}. 
\begin{figure}[h!]
	\begin{subfigure}[t]{0.5\textwidth}
		\centering
		\includegraphics[width=\linewidth]{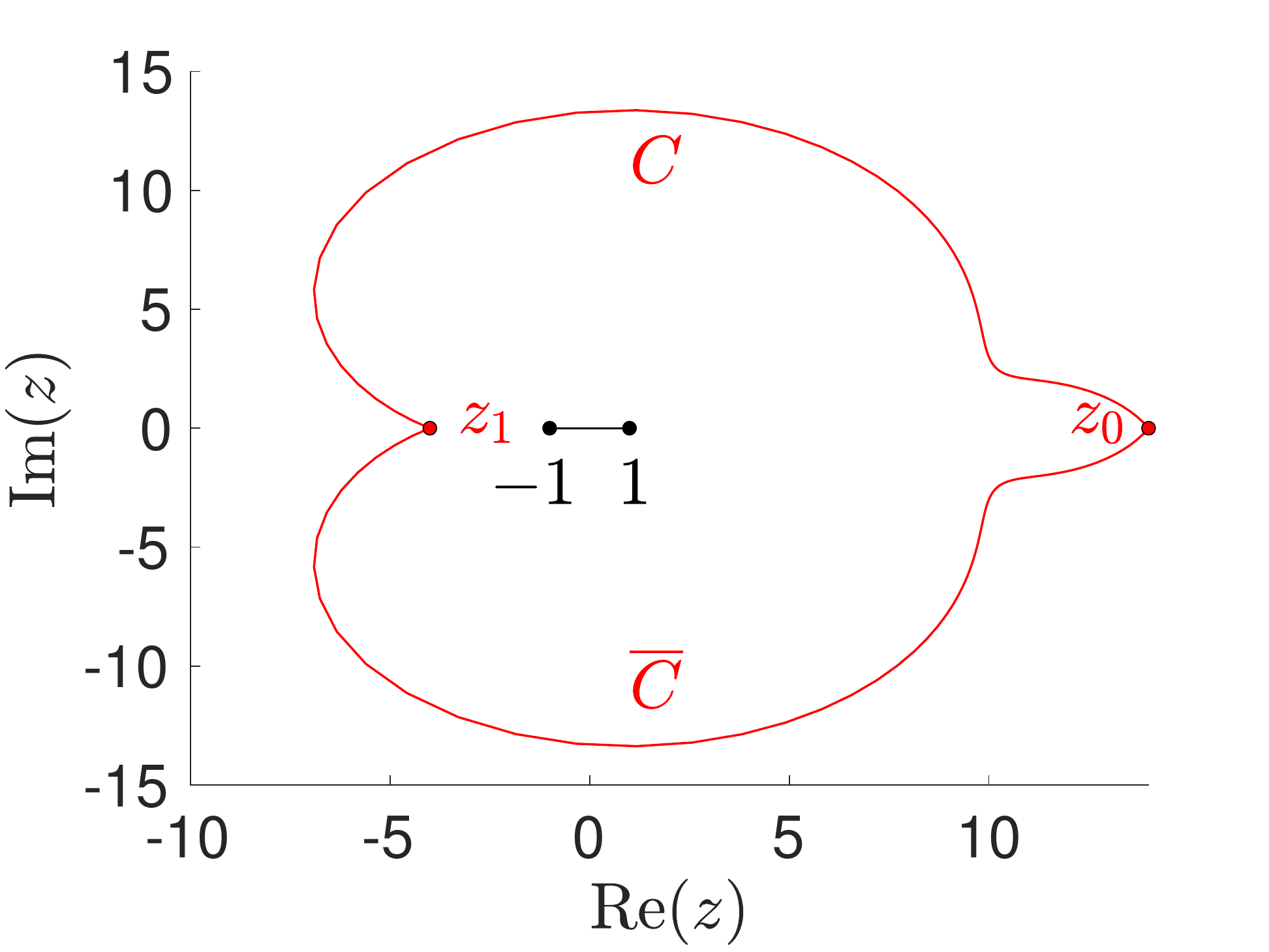}
		\caption{}
	\end{subfigure}	
	\begin{subfigure}[t]{0.5\textwidth}
		\centering
		\includegraphics[width=\linewidth]{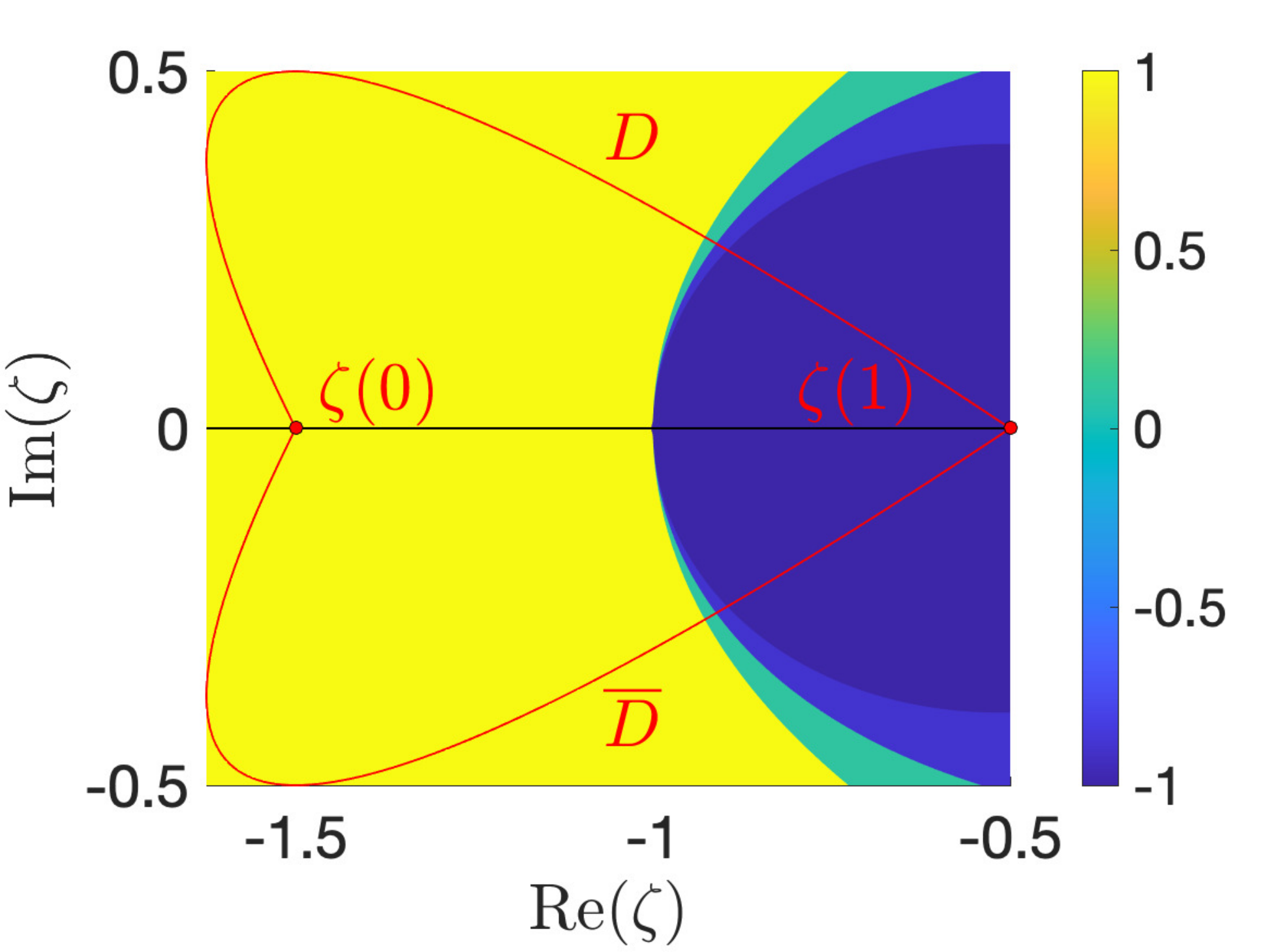}
		\caption{}
	\end{subfigure}
	\caption{(a) Given $\mu_1=1-1/\omega^2$ and $\mu_2=1+i/\omega$, we choose the trajectory $\omega(s)=-2(1+i)s^2+(2+i)s+1.5i$, with $s\in[0,1]$, so that its image through the function $z(\omega)$ given by \eqref{2.3} is the red curve connecting the points $z(\omega(0))=14$ and $z(\omega(1))=-4$. Then, $C\cup\overline{C}$ is a closed curve encircling the interval $[-1,1]$ once. (b) In red is the curve $D$ traced by  $\Gz(s)=i\omega(s)$ as $s$ is increased from 0 to 1, and the curve $\overline{D}$ traced by $\overline{\Gz(s)}$. The domain inside $D\cup\overline{D}$ is $\Omega$. In black is the straight line where the function  $h(\Gz)=z(-i\Gz)$  takes real values, the yellow region is where the real part of $h(\Gz)$ takes values bigger than 1 and  the purple region where it takes values smaller than $-1$. The color bar indicates that the real part of $h(\Gz)$ never takes values $[-1,1]$ in the domain $\Omega$. Notice that \cb $h(\Gz)$ \cb has a single pole at $\zeta=-1$: indeed, $z(\omega)$ has a pole at $\omega_*=i$.}
	\labfig{fig:curves_model2}
\end{figure}

\begin{figure}[h!t]
	\begin{subfigure}[t]{0.5\textwidth}
		\centering
		\includegraphics[width=\linewidth]{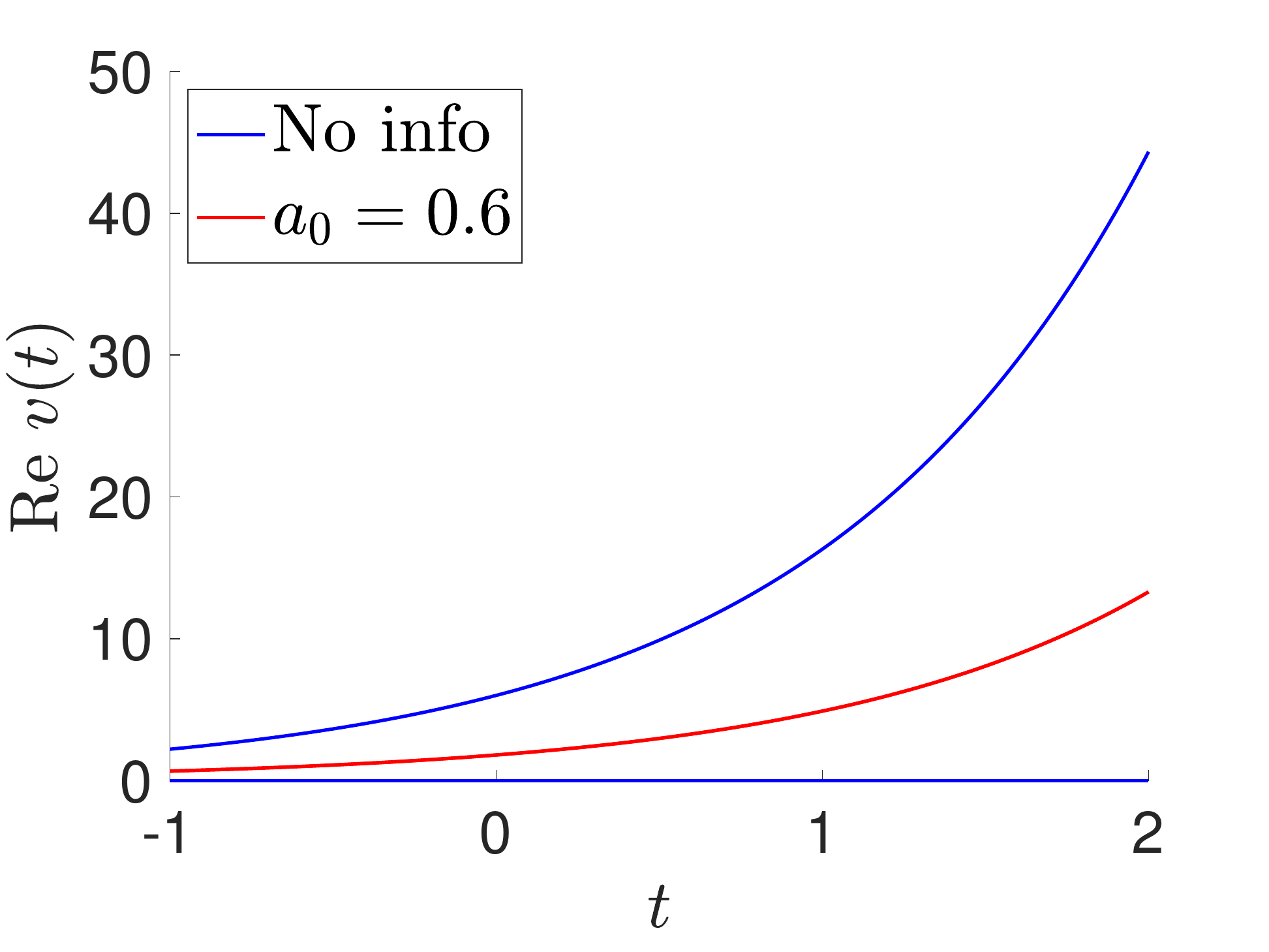}
		\caption{}
	\end{subfigure}	
	\begin{subfigure}[t]{0.5\textwidth}
		\centering
		\includegraphics[width=\linewidth]{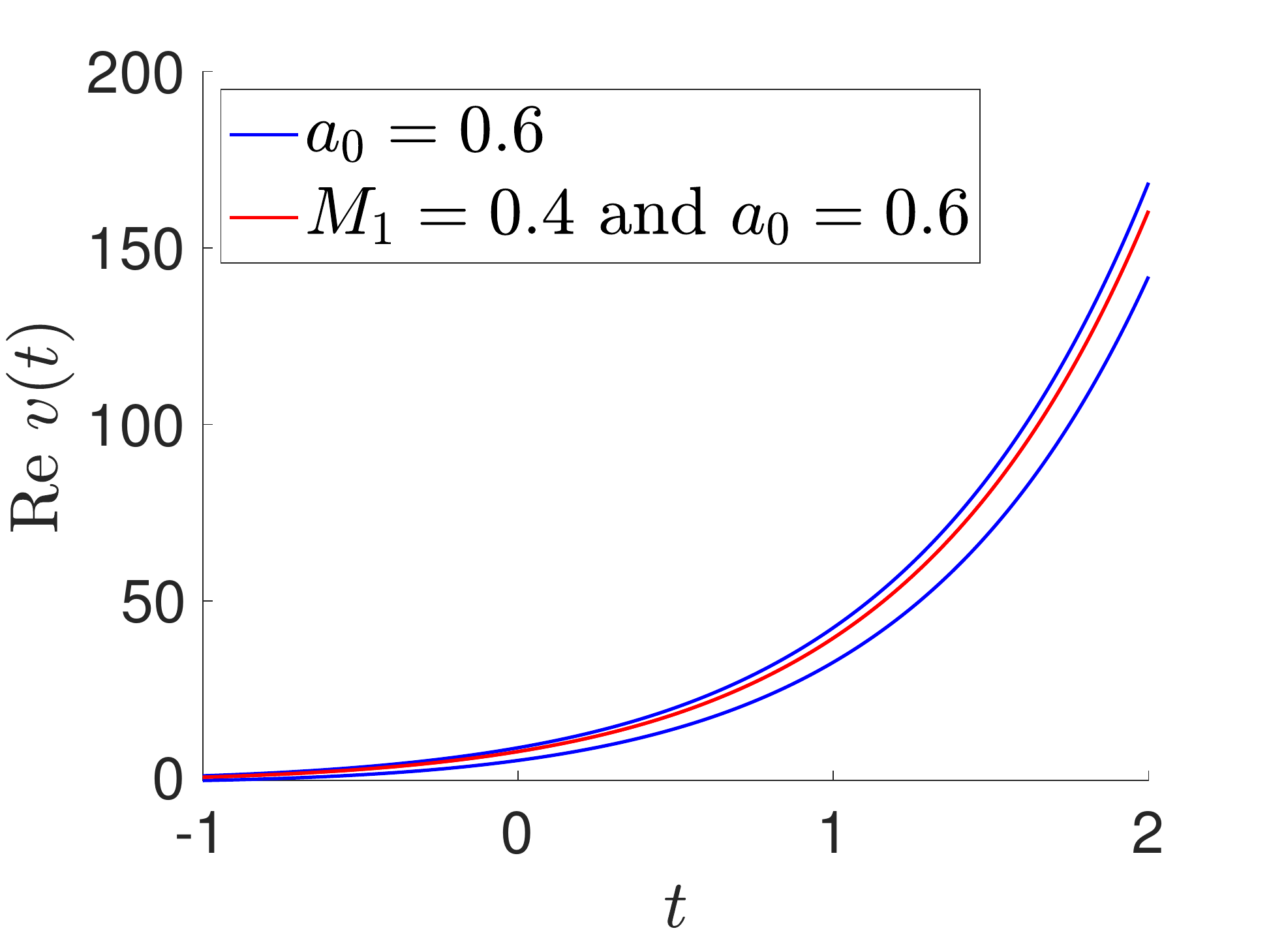}
		\caption{}
	\end{subfigure}
	\caption{The case where $\mu_1=1-1/\omega^2$, $\mu_2=1+i/\omega$, and $\omega(s)=-2(1+i)s^2+(2+i)s+1.5i$. (a) Bounds on the output function $\Real[v(t)]$: in blue are the bounds when the volume fraction, and therefore $a_0$, is not given, whereas in red are the bounds incorporating the volume fraction $f_1$ ($a_0=2f_1=0.6$). The latter bounds are coincident: indeed, the spectrum of frequencies $\omega(s)$ chosen is such that the bounds are measure-independent at any moment of time. Their analytical expression is exactly determined by \eqref{4.12}. (b) Bounds on $\Real[v(t)]$ when only the the volume fraction $f_1$ ($a_0=2f_1=0.6$) is given (outer blue bounds), whereas in red are the bounds incorporating also the first moment of the measure $M_1=0.4$. Again, the latter bounds  are coincident: like in the case when the first moment of the measure was not known, the spectrum of frequencies $\omega(s)$ chosen is such that the bounds are measure-independent at any moment of time. 
		Furthermore, their analytical expression is exactly determined by the sum of equations \eqref{4.12} and \eqref{4.25}.}
	\labfig{fig:bounds_Re_v2}
\end{figure}

Finally, we engineer an example in which the trajectory $\omega(s)$ is such that the bounds are measure dependent at any $t$, except at $t=t_0=0$, where they are coincident. To this purpose, consider a material for which $z(\omega)=(i\omega+\alpha_2)/(\omega^2-i(\alpha_1+\alpha_3)\omega-\alpha_1\alpha_3)$, with $\alpha_1<\alpha_2<\alpha_3$ real coefficients, and the following trajectory $\omega(s)=(\alpha_1+\epsilon)i+bs-(b+(2\epsilon +\alpha_1-\alpha_3)i)s^2$, where $\epsilon$ is a sufficiently small real number and $b$ is possibly complex. By choosing the parameters wisely, as in \protect{\fig{fig:curves_not_coin}(a)}, the closed curve $C\cup\overline{C}$ traced by $z(\omega(s))$ and $\overline{z(\omega(s))}$, as $s$ varies in $[0,1]$, encircles the interval $[-1,1]$, while the function $h(\Gz)=z(-i\Gz)$ takes real values between -1 and 1 in the domain $\Omega$, see \protect{\fig{fig:curves_not_coin}(b)}.  As a result, the bounds on the output function are coincident only at $t=t_0=0$, as shown in \protect{\fig{fig:bounds_not_coinc}}.
\begin{figure}[h!]
	\begin{subfigure}[t]{0.5\textwidth}
		\centering
		\includegraphics[width=\linewidth]{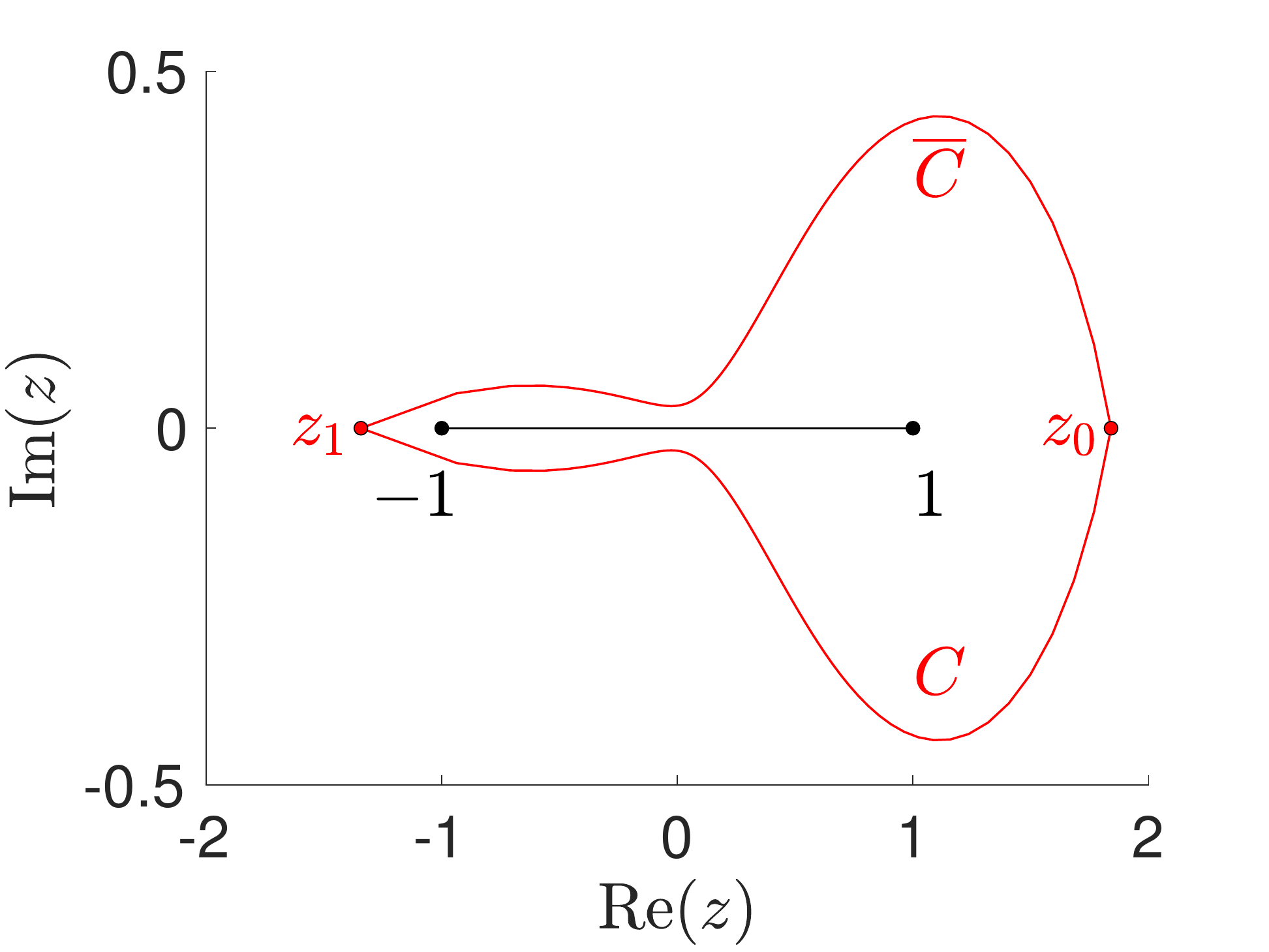}
		\caption{}
	\end{subfigure}	
	\begin{subfigure}[t]{0.5\textwidth}
		\centering
		\includegraphics[width=\linewidth]{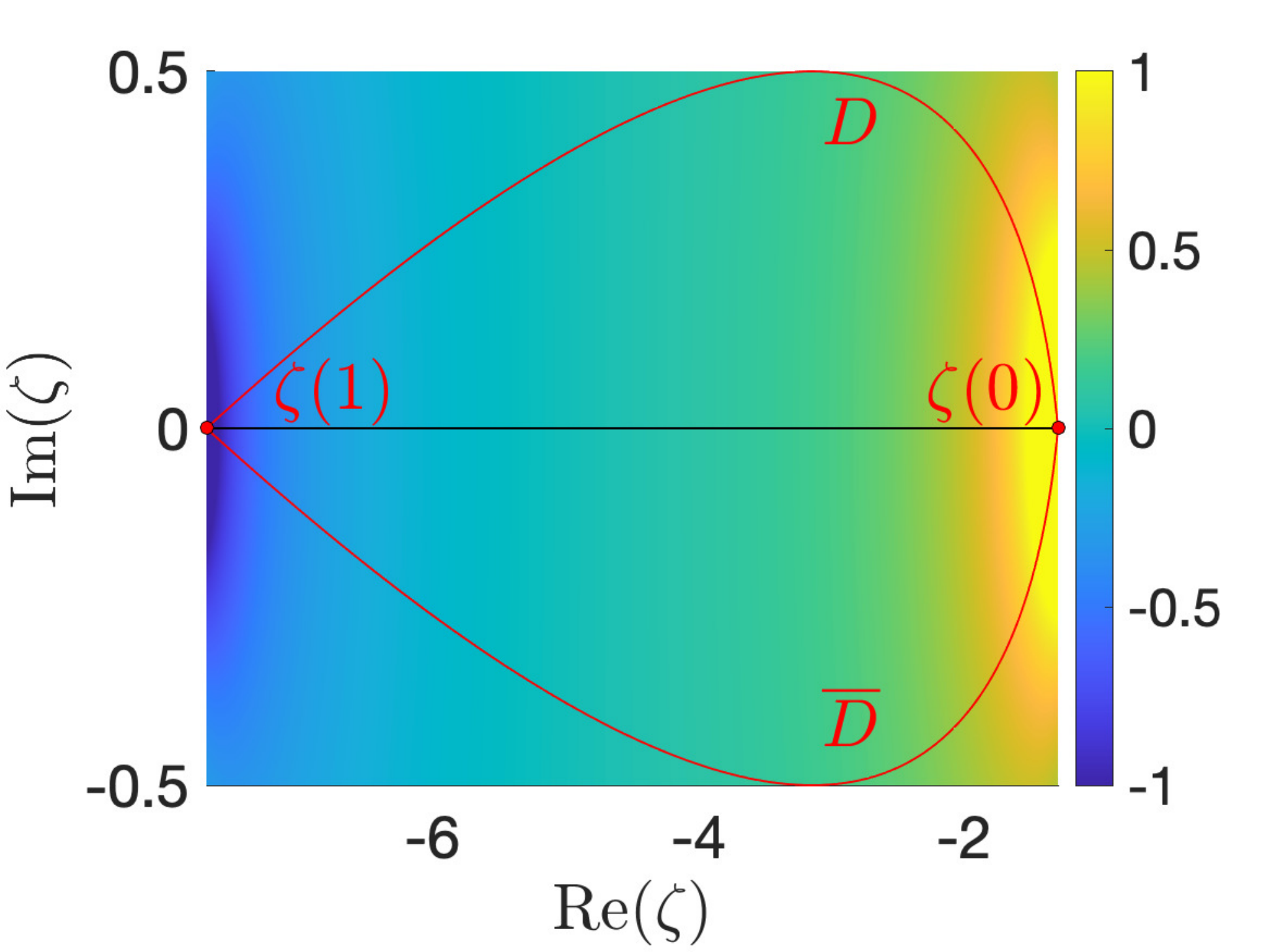}
		\caption{}
	\end{subfigure}
	\caption{(a) Given $z(\omega)=(i\omega+\alpha_2)/(\omega^2-i(\alpha_1+\alpha_3)\omega-\alpha_1\alpha_3)$, with $\alpha_1=1$, $\alpha_2=5$, and $\alpha_3=8$, we choose the trajectory $\omega(s)=(\alpha_1+\epsilon)i+bs-(b+(2\epsilon +\alpha_1-\alpha_3)i)s^2$, with $\epsilon=0.3$ and $b=2+i$, with $s\in[0,1]$, so that its image through the function $z(\omega)$ given by \eqref{2.3} is the red curve connecting the points $z(\omega(0))=1.8408$ and $z(\omega(1))=-1.3433$. Then, $C\cup\overline{C}$ is a closed curve encircling the interval $[-1,1]$ once. (b) In red is the curve $D$ traced by  $\Gz(s)=i\omega(s)$ as $s$ is increased from 0 to 1, and the curve $\overline{D}$ traced by $\overline{\Gz(s)}$. The domain inside $D\cup\overline{D}$ is $\Omega$. In black is the straight line where the function  $h(\Gz)=z(-i\Gz)$ is real. The choice of  parameters, for which $h(\Gz)$  has two single poles at $\zeta=-8$ and at $\zeta=-1$, while it has a zero at $\zeta=-5$, is such that $h(\Gz)$  does take real values between $-1$ and $1$ in the domain $\Omega$, as indicated by the color bar.}
	\labfig{fig:curves_not_coin}
\end{figure}
\begin{figure}[h!t]
	\begin{subfigure}[t]{0.5\textwidth}
		\centering
		\includegraphics[width=\linewidth]{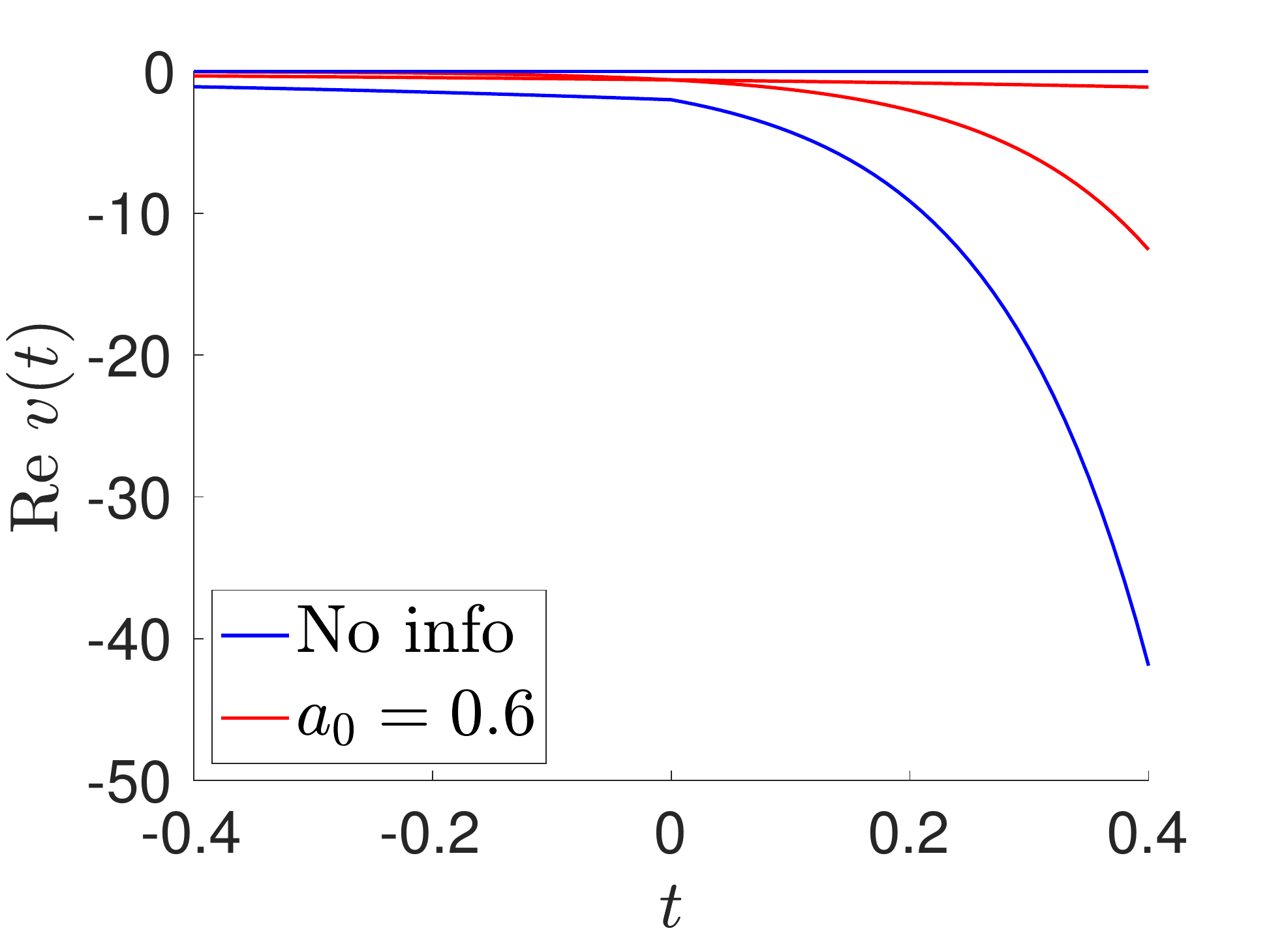}
		\caption{}
	\end{subfigure}	
	\begin{subfigure}[t]{0.5\textwidth}
		\centering
		\includegraphics[width=\linewidth]{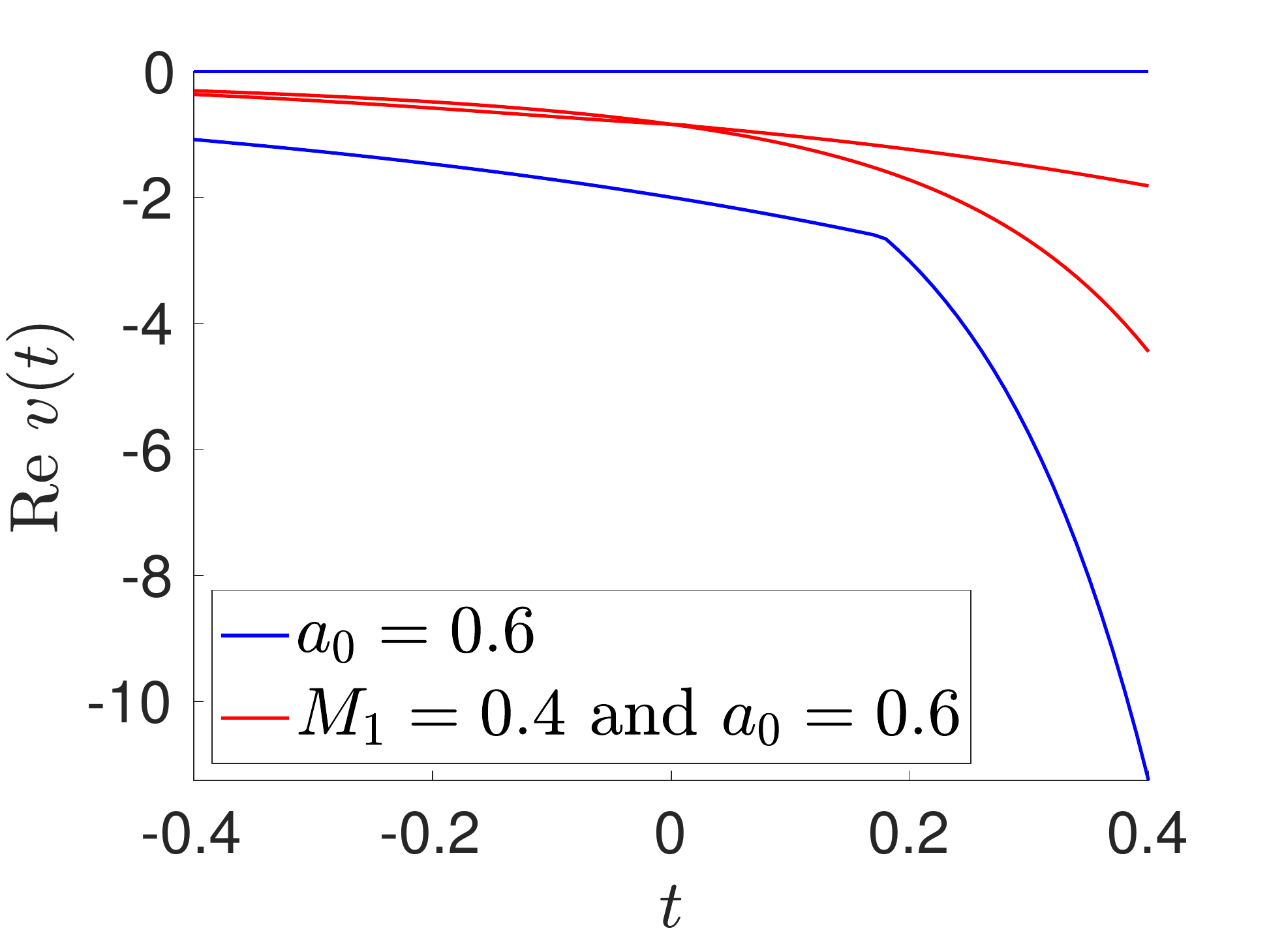}
		\caption{}
	\end{subfigure}
	\caption{Consider $z(\omega)=(i\omega+\alpha_2)/(\omega^2-i(\alpha_1+\alpha_3)\omega-\alpha_1\alpha_3)$, with $\alpha_1=1$, $\alpha_2=5$, and $\alpha_3=8$, and trajectory $\omega(s)=(\alpha_1+\epsilon)i+bs-(b+(2\epsilon +\alpha_1-\alpha_3)i)s^2$, with $\epsilon=0.3$ and $b=2+i$, with $s\in[0,1]$. (a) Bounds on the output function $\Real[v(t)]$: in blue are the bounds when the volume fraction, and therefore $a_0$, is not given, whereas in red are the bounds incorporating the volume fraction $f_1$ ($a_0=2f_1=0.6$). Contrariwise to the previous cases, the latter bounds are coincident only at $t=t_0=0$, where they take value $-a_0$: indeed, the spectrum of frequencies $\omega(s)$ chosen is such that the bounds are measure-dependent at any moment of time (except $t=t_0$). (b) Bounds on $\Real[v(t)]$ when only the the volume fraction $f_1$ ($a_0=2f_1=0.6$) is given (outer blue bounds), whereas in red are the bounds incorporating also the first moment of the measure $M_1=0.4$. Again, the latter bounds  are coincident only at time $t=t_0=0$.}
	\labfig{fig:bounds_not_coinc}
\end{figure}

\section{Conclusions}

{In this paper, we consider the response in time of a two-phase composite in which at least one of the two constituent materials  has a non-local response in time. We found that when the applied field is suitably chosen, the measurement of the response of the composite at a specific moment of time exactly yields the volume fraction of the phases, independent of the microstructure of the composite. To achieve such a result, one has to choose affine boundary conditions where the applied field is composed by a continuous spectrum of complex frequencies. Such spectrum has to satisfy certain properties that are related to the response of the constituent materials, which are supposed to be rational functions of the frequency. This is to guarantee that there exists a range of frequencies where the absorption is zero (to this regard, see for instance the experiment described in \cite{Lakes:1996:VBI}, where the authors studied viscous damping in a material over 7 decades of frequency). If that were not the case, then one could expect the trajectory of frequencies to fail to satisfy some of the properties that would ensure the exact determination of the volume fraction (specifically, the curve $C$ would not cross the horizontal axis and, therefore, $C\bigcup \overline{C}$ would not form a closed loop around the interval $[-1,1]$).
If the continuous spectrum of frequencies satisfies the further property that the initial and final frequencies are purely imaginary, then, remarkably, the volume fraction can be retrieved by measuring the response of the composite at any moment of time. Note that a  slight modification of the boundary conditions will also allow one to recover the first moment of the measure. Additionally, one can recover the response of the composite at a certain frequency by suitably applying boundary conditions not oscillating at such a frequency. }{Finally, following \cite{Mattei:2020:EPA}, all the analysis developed here
carries through to determining exactly the Fourier components of an inclusion in a body, from which the shape of the inclusion and not just its size
can theoretically be recovered.}

{Note that our results assume the absence of noise and that the responses
of the phases are exactly known. Noise in the measurements of the output at
a given time, say $t = t_0$ is easily handled. Different volume fractions would
correspond to a different output at time $t=t_0$. Those outputs
compatible with the error bars of the measured response at $t=t_0$ give us the range of possible
volume fractions. Dealing with uncertainity in the responses
of the phases is more involved. 
The functions $\Gm_1(\omega)$ and $\Gm_2(\omega)$ need to have the required
analytic properties. So, instead of these
being exactly known, they may belong to sets of functions $\BGY_1$ and $\BGY_2$ satisfying these properties.
Then for every pair of functions, one in each set, associated bounds need to be 
calculated on the volume fraction from the measured response at time $t=t_0$. However, the input function
should be that for one specific pair and such that the volume
fraction would be uniquely determined by measurements at $t = t_0$ were this
pair the actual responses of the two phases. In the end, one should take the union of these bounds on the volume
fraction. In addition to the complexity, the effect of incorporating a
range of functions $\Gm_1(\omega)$ and $\Gm_2(\omega)$ would likely be highly dependent on the
chosen example. So, faced with a particular problem it would be prudent for
a researcher to investigate these effects.}

\section*{Acknowledgment}

The authors are grateful to the National Science Foundation for support through
Research Grants DMS-1814854, DMS-2107926, and DMS-2008105. M.P. was partially supported by a Simons Foundation collaboration grant for mathematicians. Davit
Harutyunyan is thanked for his help in the initial stages of this work. {The authors also wish to thank the reviewers for their insightful comments and suggestions that helped improved the manuscript}.

\ifx \bblindex \undefined \def \bblindex #1{} \fi\ifx \bbljournal \undefined
  \def \bbljournal #1{{\em #1}\index{#1@{\em #1}}} \fi\ifx \bblnumber
  \undefined \def \bblnumber #1{{\bf #1}} \fi\ifx \bblvolume \undefined \def
  \bblvolume #1{{\bf #1}} \fi\ifx \noopsort \undefined \def \noopsort #1{}
  \fi\ifx \bblindex \undefined \def \bblindex #1{} \fi\ifx \bbljournal
  \undefined \def \bbljournal #1{{\em #1}\index{#1@{\em #1}}} \fi\ifx
  \bblnumber \undefined \def \bblnumber #1{{\bf #1}} \fi\ifx \bblvolume
  \undefined \def \bblvolume #1{{\bf #1}} \fi\ifx \noopsort \undefined \def
  \noopsort #1{} \fi

%
%
%

\end{document}